\documentclass[a4paper,11pt]{article}
\pdfoutput=1 

\usepackage{jheppub} 

\usepackage[T1]{fontenc} 
\usepackage{comment}
\usepackage{pdfpages}

\graphicspath{ {image/} }

\title{\boldmath Entanglement of Purification for Momentum Relaxed Superconductor}

\author[a]{I-Hsi Chen,}
\author[a]{Po-Shuo Huang,}
\author[]{and Shang-Yu Wu}

\affiliation[a]{Department of Physics, National Taiwan University,\\Taipei 10617, Taiwan}

\emailAdd{R05222043@ntu.edu.tw}
\emailAdd{ryan840919@gmail.com}
\emailAdd{loganwu@gmail.com}

\abstract{We reconstruct the information quantities in the holographic relaxed superconductor system 
and discuss how these quantities behave under non-symmetry configuration. We then combine the effect of the superconductor and the momentum relaxation, and find out that the superconductor effect will change how the momentum relaxation affect these information quantities.}

\begin{document} 
\maketitle
\flushbottom

\section{Introduction}
\qquad Quantum information has been a hot topic during these years, but sometimes it is somehow difficult to measure these information quantities from the quantum field theory perspective. Thanks to the AdS/CFT duality, we may have a chance to see these quantities from the geometric side.

There are many information quantities have been dual to the gravity side, such as entanglement entropy(EE)\cite{Ryu:2006bv}, mutual information(MI) and entanglement of purification(EoP)\cite{Takayanagi:2017knl}. These quantities have different meaning and are suit for different situation. For pure state, EE is a good measurement of entanglement, but for mixed state, classical correlation comes in so we need to consider the MI - the linear combination of EE or the Eop - the generalization of EE for mixed state, which better describes correlation.

EoP and MI have been discussed in many holographic situations, such as pure AdS or AdS-RN black hole in \cite{Liu:2019qje}. In this research, they discuss how these quantities depend on temperature and non-symmetry configuration. They have also been discussed in the holographic superconductor model in \cite{Liu:2020blk} and it shows that these information quantities have corresponding behaviors with the phase transition. In \cite{Huang:2019zph}, they used the axion model to discuss how the momentum relaxation affect these quantum information quantities. In this paper, we want to combine these discussion to a more complicated model, which is the holographic momentum relaxed superconductor system.

To construct a momentum relaxed superconductor, we need to break the translation symmetry by adding two massless real scalar fields, which are linear to spatial coordinates with proportionally constant $\beta$. Follow the steps in \cite{Kim:2015dna}, we know how this momentum relaxation affect the phase transition and other properties of the superconductor. Inspired by this, we were curious about how this momentum relaxation affect the EoP and MI in the superconductor phase, and wonder if in this situation the EoP and MI are still good measurements of the phase transition.

The paper is organized as followed: In section \ref{holographic relaxed superconductor} we reconstruct the holographic superconductor with momentum relaxation, and check our result by calculating the T to condensate behavior. In section \ref{Holographic entanglement of purification} we analyze the entanglement of purification and mutual information in holographic system. We calculate the EoP and MI in the holographic relaxed system and discuss all the effect do to these information quantities in section \ref{results}. Finally, we conclude in section \ref{conclusion}.

\section{Holographic Relaxed Superconductor}\label{holographic relaxed superconductor}

\qquad To develop a holographic superconductor system, followed \cite{Hartnoll:2008vx,Hartnoll:2008kx}, we can write down the Lagrangian density for a Maxwell field and a charged complex scalar field coupled to gravity:
\begin{equation}
    \mathcal{L}=R+\frac{6}{L^2}-\frac{1}{4}F^{ab}F_{ab}-V(|\psi|)-|\nabla\psi-iqA\psi|^2
\end{equation}
but in our case, we want to add an extra massless scalar field term to break the translation symmetry, which is first introduced in \cite{Andrade:2013gsa}:
\begin{equation}
    \mathcal{L}=-\frac{1}{2}\sum_{I=1}^{d-1}(\partial\xi_I)^2
\end{equation}
This will lead us to a holographic superconductor with momentum relaxation. We are actually follow the steps in \cite{Kim:2015dna}.

Follow the ward identity in \cite{Andrade:2013gsa}, we know that turning on sources in the presence of non-zero vevs yields the non-conservation of $\langle P^a \rangle\equiv\langle T^{ta} \rangle$, which will turn on the momentum-relaxation effect. By the (A.10) in the appendix in \cite{Kim:2015dna}, we know that in the superconductor cases the stress-energy tensor $T_{\mu\nu}$ will not explicitly depend on the coordinates of the boundary, this is due to the "massless" scalar field imply in \cite{Andrade:2013gsa}. The massless property will let the scalar field enter the stress-tensor through $\partial_{\mu}\xi$. This will make our calculation much easier, or in more detail, this will make our background homogeneous and let the EOMs all become ODE.

We are looking for electrically charged plane-symmetric hairy black hole solutions with scalar field backreact the background spacetime(\cite{Kim:2015dna}). Thus we take the metric:
\begin{equation}
    ds^2=-g(r)e^{-\chi(r)}dt^2+\frac{dr^2}{g(r)}+r^2(dx^2+dy^2)
\end{equation}
with
\begin{equation}
    A=\phi(r)dt, \quad \psi=\psi(r) , \quad \xi_I=\beta_{Ii}x^i=\frac{\beta}{L^2}\delta_{Ii}x^i
\end{equation}
by plug in this ansatz into the equation of motion, we can get:
\begin{equation}
\begin{split}
    &\psi''+(\frac{g'}{g}-\frac{\chi}{2}+\frac{2}{r})\psi'+\frac{q^2\phi^2e^\chi}{g^2}\psi-\frac{1}{2g}V'(\psi)=0\\
    &\phi''+(\frac{\chi'}{2}+\frac{2}{r})\phi'-\frac{2q^2\psi^2}{g}\phi=0\\
    &\chi'+r\psi'^2+\frac{rq^2\phi^2\psi^2e^\chi}{g^2}=0\\
    &\frac{1}{2}\psi'^2+\frac{\phi'^2e^\chi}{4g}+\frac{g'}{gr}+\frac{1}{r^2}-\frac{3}{gL^2}+\frac{V(\psi)}{2g}+\frac{q^2\psi^2\phi^2e^\chi}{2g^2}=-\frac{\beta^2}{2r^2gL^2}
\end{split}
\end{equation}
which we already set our $\psi$ to be real, and potential $V(\psi)=-2\psi^2/L^2$,
and in this part we know that the Hawking temperature of the black hole is:
\begin{equation}
\begin{split}
    &T=\frac{g'(r_h)e^{-\frac{1}{2}\chi(r_h)}}{4\pi}
    \\
    &=\frac{r_h}{16\pi L^2}(12-2m^2L^2\Psi(r_h)^2-2\frac{\beta^2}{r_h^2}-L^2e^{\chi(r_h)}(\phi'(r_h))^2)e^{-\frac{1}{2}\chi(r_h)}
\end{split}
\end{equation}
We want to emphasize that by the dimensional analysis,
\begin{equation}
    [T]=M, \quad [\mu]=M, \quad [\psi^{(2)}]=M^2, \quad [\beta]=M
\end{equation}
To achieve the dimensionless quantities, we will set our $\Tilde{T}$ = T/$\mu$ and $\tilde{\psi^{(2)}}$ = $\psi^{(2)}/\mu^2$ to calculate the condensate with temperature, but abandon the "tilde" notation for convenience, and to see how the momentum relaxation actually affect the superconductor properties, instead of consider each $\beta$ cases, followed \cite{Kim:2015dna}, we reconstruct different ratio of $\beta/\mu$ of superconductors, which all of them are dimensionless quantity, we will mention this again in Sec~\ref{results}.

Our condition on the boundary side is $\psi^{(1)}$=0, and we can use our three parameters $\psi(r_h)$, $\phi'(r_h)$ and $\chi(r_h)$ to shoot from horizon by fixing every $\beta$ and try to find the right $\mu$ for each $\beta$.

To let our black hole temperature equal to the boundary CFT temperature, we need to set our $\chi$ to zero at the boundary, which will achieve the asymptotic AdS space, in other words:
\begin{equation}
    \chi \to 0,\quad r\to\infty 
\end{equation}
we can use one of the symmetry of equation of motions:
\begin{equation}
    e^{\chi}\to a_1^2e^{\chi},\quad t\to a_1t,\quad \phi\to \phi/a_1
\end{equation}
to set the $\chi$($\infty$)=0, just by setting $a_1$=$e^{-\chi(0)/2}$. Keep in mind that after using this symmetry, we also need to divide our $\mu$ and T a ($e^{-\chi(0)/2}$) coefficient, so we can get the $\mu$ and T for the asymptotic AdS situation. For example, when we mentioned we shoot from horizon with fixing $\beta$ and try to find the right $\mu$ for each $\beta$, we are actually try to find the right $\mu$ and $\chi(0)$ combination($\mu e^{\chi(0)/2}$) to match the right ratio for each $\beta$. 

There are also other two symmetries of equation of motions:
\begin{equation}
\begin{split}
    &r\to a_2r,\quad (t,x,y)\to a_2(t,x,y) ,\quad  L\to a_2L ,\quad q\to q/a_2, \quad m\to m/a_2, \quad \beta\to a_2\beta\\
    &r\to a_3r, \quad (t,x,y)\to (t,x,y)/a_3, \quad g\to a_3^2g, \quad \phi\to a_3\phi, \quad \beta\to a_3\beta
\end{split}
\end{equation}
with these symmetries, we can scale our L and $r_h$ to 1, which will take a big advantage of simplify the numerical calculation.

After all the numerical calculation above, we plot the T to condensate to show that it fits the superconductor behavior, and it agreed with the result in \cite{Kim:2015dna}.

\begin{figure}[h!]
    \centering
    \includegraphics[scale=0.7]{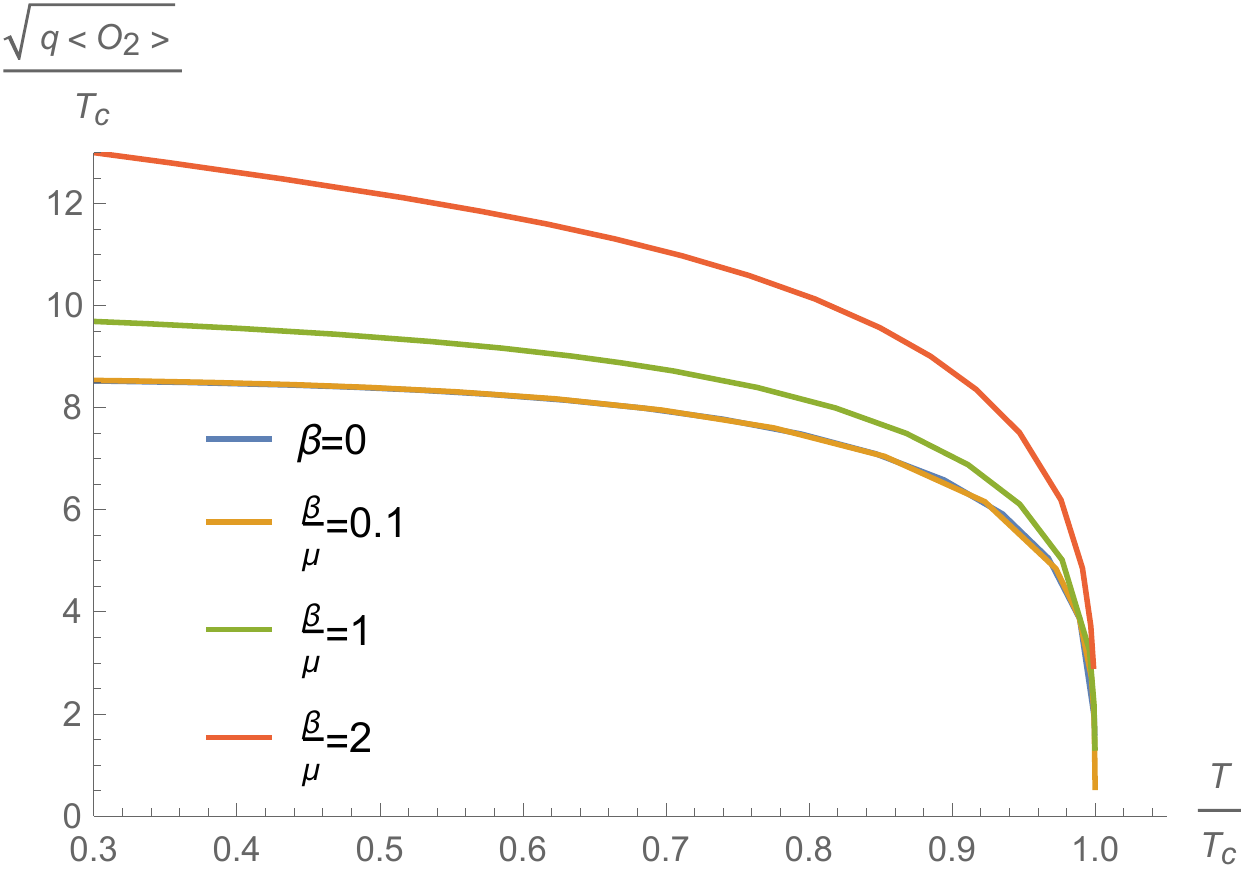}
    \caption{T vs. condensate with different $\beta/\mu$ ratio.}
    \label{fig:condensate to T}
\end{figure}

\section{Holographic entanglement of purification}
\label{Holographic entanglement of purification}
\begin{figure}[h!]
    \centering
    \includegraphics[scale=0.7]{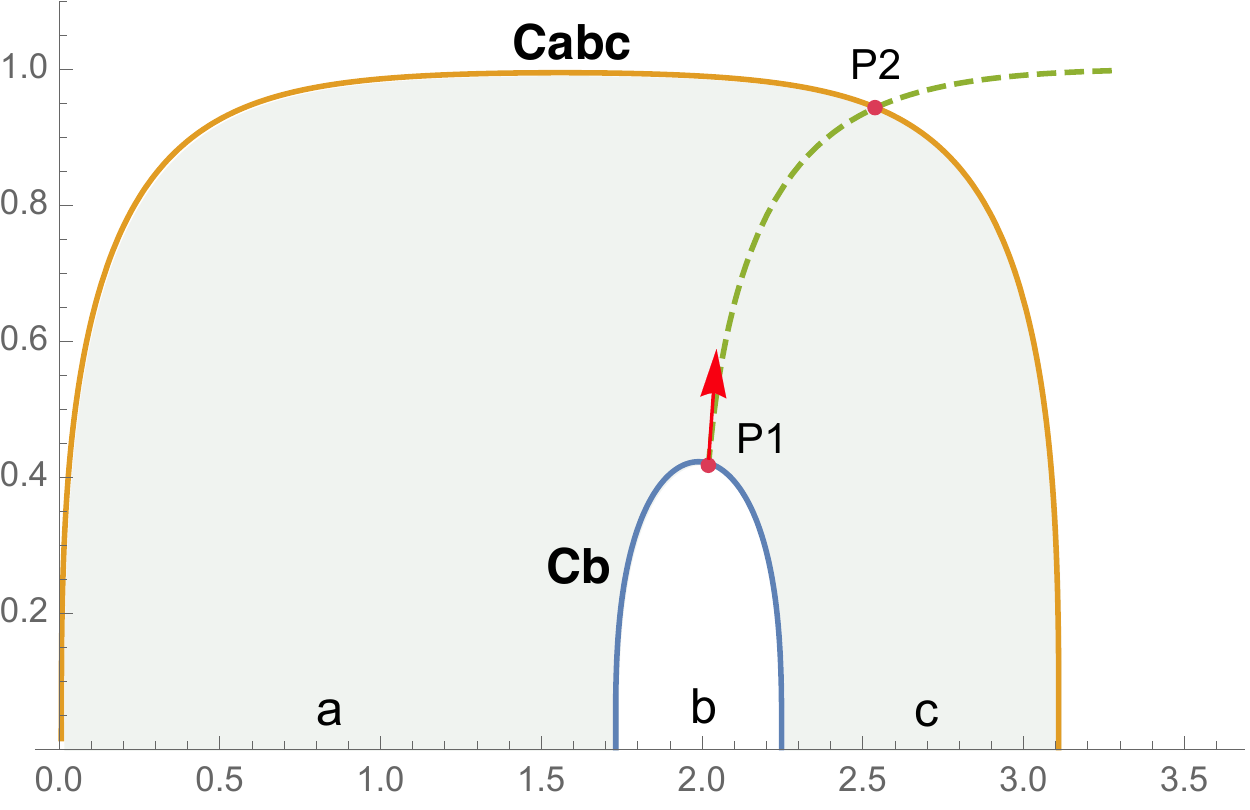}
    \caption{The entanglement wedge is the area between blue and yellow line. The EoP corresponds to the minimum cross-section of entanglement wedge, which is proportional to the arc-length of the geodesic(green dashed line) connecting $P_1$ and $P_2$ .}
    \label{fig:entanglement wedge}
\end{figure}

\qquad A well known relation between entropy and area is the Bekenstein-Hawking entropy for black holes\cite{Bekenstein:1973ur,Hawking:1974sw}. The entropy is not proportional to its volume but its surface area of the horizon, which manifest the holographic description of the AdS/CFT correspondence that the quantum gravity on $(d+2)$-dimensional anti-de Sitter
spacetime is equivalent to a conformal field theory in $(d+1)$ dimensions. This indicates that the information is recorded on the $(d+1)$ dimensional boundary.
In the work of Ryu and Takayanagi \cite{Ryu:2006bv} the entanglement entropy is proposed to follow the area law
\begin{equation}
S_A=\frac{Area(\gamma_A)}{4G_N^{(d+2)}}
\end{equation}
where $\gamma_A$  is the $d$-dimensional minimal surface whose boundary is given by the $(d-1)$-
dimensional manifold $\partial\gamma_A$.
They give a simple calculation  showing that entanglement entropy in AdS$_3$ agrees with the result in the $(1+1)$D CFT side\cite{Holzhey:1994we}
\begin{equation}
S_A=\frac{c}{3}\log\frac{l}{a}
\end{equation}
where c is the central charge of the CFT.
With this conjecture, it was deduced that the entanglement entropy of which minimal surface is deformed by black hole or other fields can be calculated in the same manner.
The Entanglement of Purification(EoP) is defined to measure correlation between two subsystems for a mixed state\cite{Terhal:2002tep}. We add an auxiliary subsystem to make the state pure then calculate its entanglement entropy. In the further work of  Takayanagi, the EoP was associated with the area of the minimum cross-section of the entanglement wedge\cite{Takayanagi:2017knl}. For the symmetric stripe configuration, the EoP is simply proportional to the arc length. Our goal is to find the geodesic which gives the minimum integral value from $P_1$ to $P_2$ along the green line, which is shown in Fig~\ref{fig:entanglement wedge}.

To construct a striped minimal surface, we start from the metric of bulk gravity. The generic line element in the bulk is
\begin{equation}
    ds^2=g_{tt}dt^2+g_{rr}dr^2+g_{xx}dx^2+g_{yy}dy^2
\end{equation}
Here we consider a stripe configuration, which means the metric is homogeneous in y direction. Taking a time slice, the minimal surface
\begin{equation}
    A=L_y\int\sqrt{g_{yy}(g_{xx}x'(s)^2+g_{rr}r'(s)^2)} ds
\end{equation}
which is parameterized by arc length and $\omega$ is stripe width. The integral can be rewrite as
\begin{equation}
\label{minimal surface 1}
    A=\int_0^{\omega}\sqrt{g_{yy}(g_{xx}+g_{rr}r'(x)^2)} dx
\end{equation}
Taking this integral as the geodesic action, the Euler-Lagrange equation gives the geodesic equation.

Now we go to the momentum relaxed superconductor metric on a constant time slice:
\begin{equation}
    ds^2=\frac{dr^2}{g_r(r)}+\frac{r^2}{L^2}(dx^2+dy^2)
\end{equation}
Making a change of variable $r=1/z$ and set $L=1$:
\begin{equation}
\label{z metric}
    ds^2=\frac{dz^2}{z^4g(z)}+\frac{1}{z^2}(dx^2+dy^2)
\end{equation}
where $g(z)$ is from changing variable of $g_r(r)$, the integral in \eqref{minimal surface 1} becomes:
\begin{equation}
\label{general minimum surface x}
    A=\frac{1}{z^2}\sqrt{1+\frac{z'(x)^2}{z^2g(z) }}dx
\end{equation}
and the EOM for the minimal surface is:
\begin{equation}
\label{mms num}
    4z^2g(z)^2 - zg'(z)z'(x)^2 + 2g(z)(z'(x)^2+z(x)z''(x))=0
\end{equation}

For $T>T_c$, the system is in normal phase, its analytical solution is
\begin{equation}
\label{normal phase metric 1}
    g(z)=\frac{1}{z^2}-\frac{\beta^2}{2}-m_0z+\frac{\mu^2z^2}{4z_h^2}
\end{equation}
where $z_h$ is the horizon in $z$ coordinate. By the condition $g(z)=0$ at the horizon, $m_0$ can be solved as
\begin{equation}
    m_0=\frac{1}{z_h^3}+\frac{\mu^2}{4z_h}-\frac{\beta^2}{2z_h}
\end{equation}
and the temperature of the black hole is:
\begin{equation}
    T=\frac{1}{4\pi}(\frac{3}{z_h}-\frac{\mu^2z_h}{4}+\frac{\beta^2z_h}{2})
\end{equation}
setting $z_h=1$ and putting $g(z)$ into \eqref{mms num}, the EOM for the minimal surface in normal phase is:
\begin{equation}
\begin{split}
    &(z-1)^2(-4+z(-4+z(-4+2\beta^2+\mu^2z)))^2+(8+z^3(4-2\beta^2+\mu^2-2\mu^2z))z'\\
    &+2(z-1)(-4+z(-4+z(-4+2\beta^2+\mu^2z)))(z'^2+zz'')=0
\end{split}
\end{equation}
when $\beta=0$ the system goes back to the AdS-RN black hole solution.

First we take the turning point of the geodesic as the boundary condition. Each $z$ coordinate of the turning point corresponds to a stripe width. The geodesic equation is translation invariant in $x$ direction, by tuning $(x,z)$ coordinate of the turning points, we are able to set the $C_b$ and $C_{abc}$ to a specific configuration and search the minimal cross-section over the possible space by moving $P_1$ along $C_b$ and tuning the shooting angle. We also use another simple technique in \cite{Liu:2020blk}, which claimed that the global minimum must also be the local minimum, follow this rule, we can find the minimum cross-section easily.

In this holographic dual, the EoP must also satisfy some inequalities of certain entanglement measures. One important measure is the mutual information. In a bipartie system the MI is defined as 
\begin{equation}
    MI(A:B):=S(A)+S(B)-S(A,B)
\end{equation}
In holographic sense, the MI vanishes when subsystems A and B become disconnected. The EoP is defined as the minimum cross-section of the entanglement wedge, so EoP vanishes when MI vanishes.
For non-vanishing MI, the inequality $E_W(\rho_{AB})\geq MI(A:B)/2$ holds, we will check it in our result.

\section{Results}
\label{results}

\qquad Followed the discussion mentioned in Sec~\ref{holographic relaxed superconductor}, all the physical parameters we discuss should be dimensionless, so after the dimension analysis, we will have:
\begin{equation}
\label{dimensionless quantity}
    \Tilde{T}=\frac{T}{\mu},\quad (\Tilde{a},\Tilde{b},\Tilde{c})=(a\mu,b\mu,c\mu),\quad \Tilde{MI}=\frac{MI}{\mu},\quad \Tilde{EoP}=\frac{EoP}{\mu}
\end{equation}
But as mentioned in Sec~\ref{holographic relaxed superconductor}, we will abandon the "tilde" sign and just write down the usual notation.

In general, we expect all cases which include the superconductor phase will appear a kink point at the critical temperature, so that it would match the non-relaxed superconductor case in \cite{Liu:2020blk} that both the EoP and MI can diagnose the phase transition in a superconductor system, and indeed in our cases they all appear. But apart from checking this, we also have done some numerical check to see our method is reliable. In Sec~\ref{consistent check},\ref{high-temperature behavior} and \ref{(a,b,c) configuration}, we will take $\beta/\mu=1$ to demonstrate the high-temperature behavior and discuss the (a,b,c) configuration effect, and in Sec~\ref{beta Effect} we will discuss how the momentum relaxation affect the EoP and MI.

\subsection{Consistency Check}
\label{consistent check}
\subsubsection{Symmetric limit}
\begin{figure}[h!]
    \centering
    \includegraphics[scale=0.42]{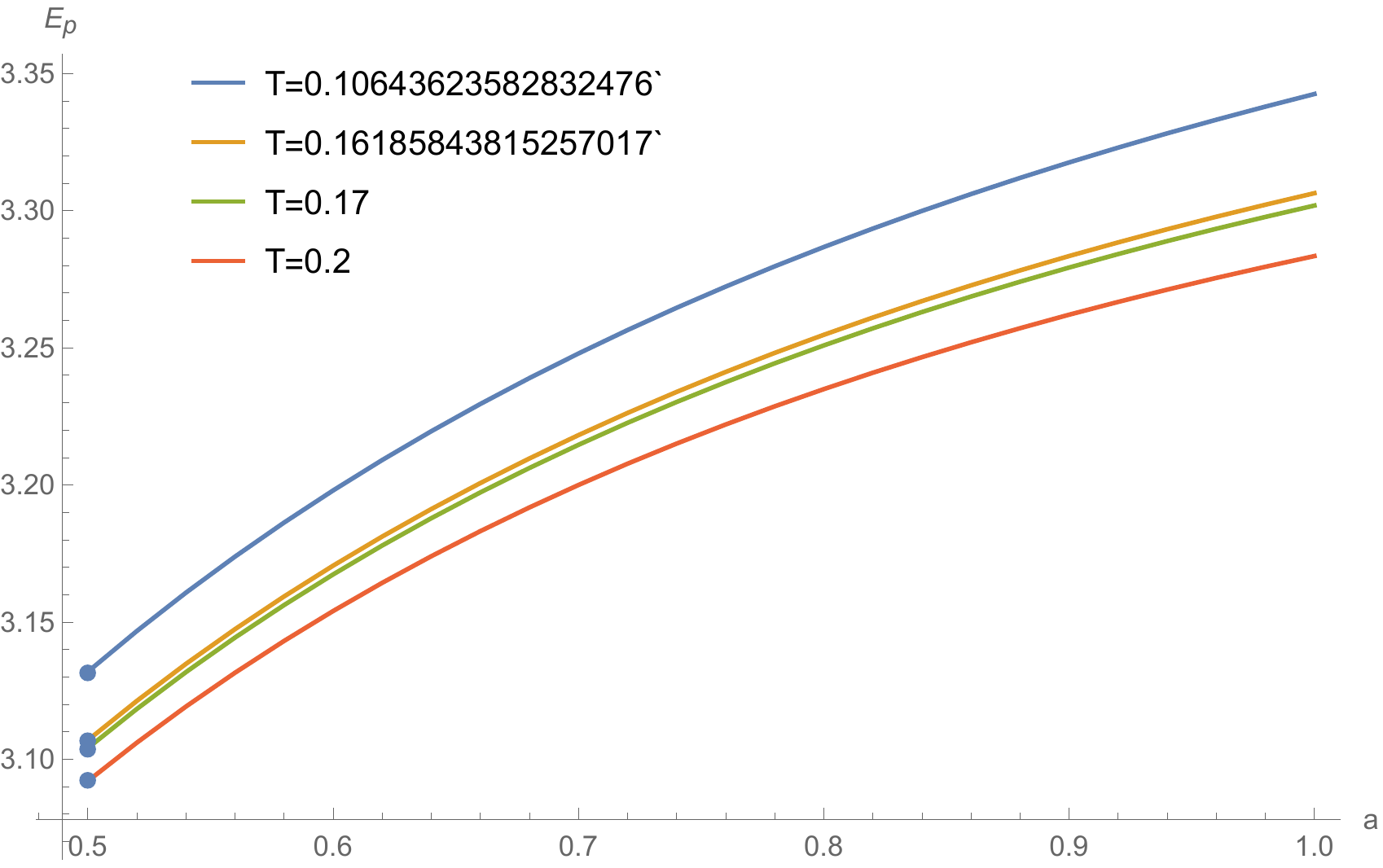}
    \includegraphics[scale=0.42]{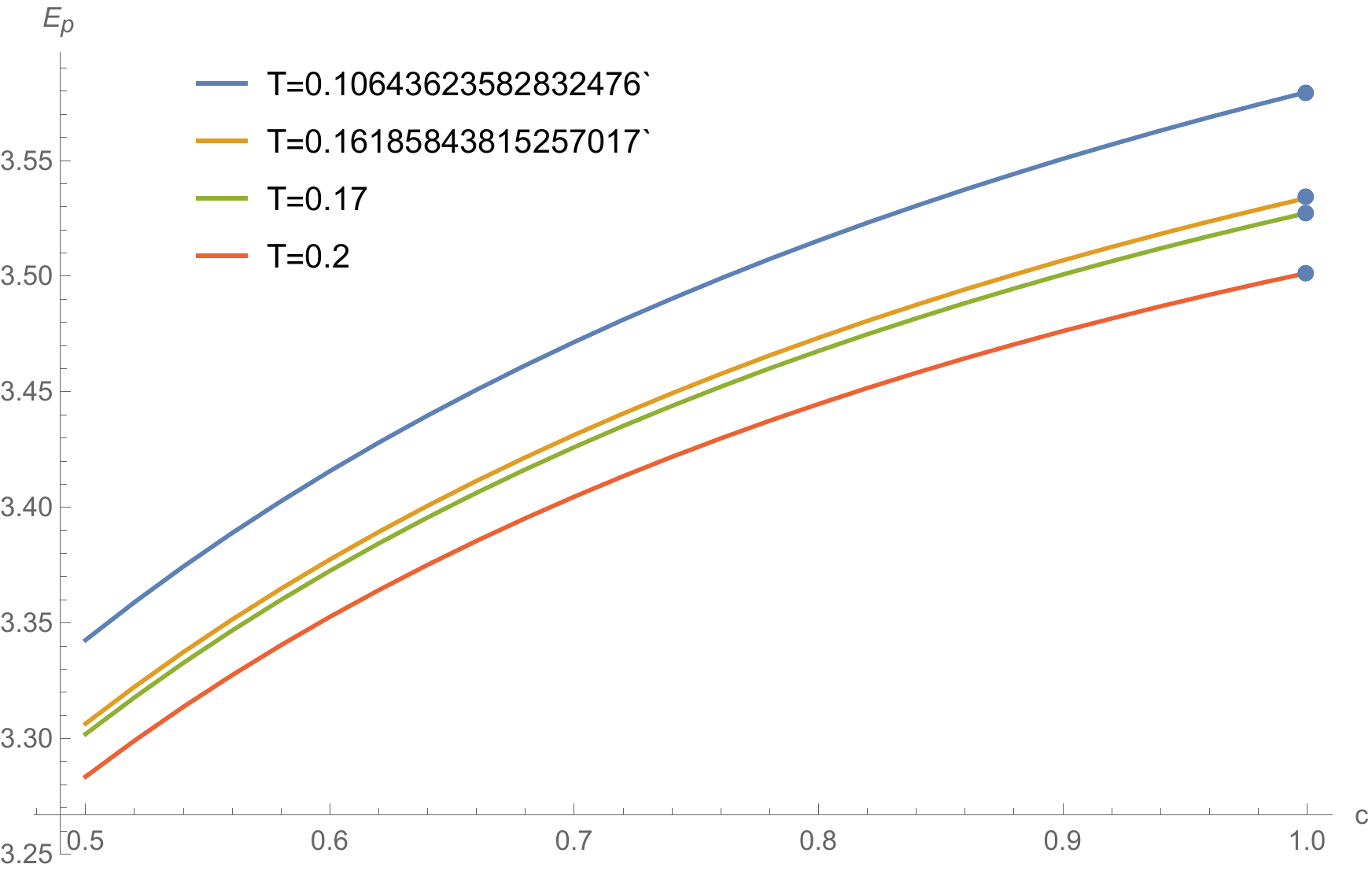}
    \caption{EoP in both symmetry and non-symmetry cases. The EoP in non-symmetry cases will approach to the symmetry cases as the configuration approach to symmetry. The left plot is from (1,0.3,0.5) approach to (0.5,0.3,0.5). The right plot is from (0,5,0.3,1) approach to (1,0.3,1). For T=0.10643623582832476` and T=0.16185843815257017`, we use the numerical metric to calculate the EoP, and for T=0.17 and T=0.2, we use the analytic metric to calculate the EoP.}
    \label{symmetry check test}
\end{figure}

By \eqref{minimal surface 1}, we know our minimum surface can also be written as
\begin{equation}
    \int_0^{\omega}\sqrt{g_{yy}(g_{xx}dx^2+g_{zz}dz^2)}
\end{equation}
for symmetry case, $dx^2$ will become zero, so we will have:
\begin{equation}
\label{minimal surface 2}
    \int_{z_1}^{z_2}\sqrt{g_{yy}(g_{zz}dz^2)}
\end{equation}
which $z_1$ and $z_2$ stand for the intersections of the $C_{b}$ and $C_{abc}$ with the minimum surface. For T<$T_c$, we can plug our numerical metric in \eqref{minimal surface 2} and get the dimensional EoP. For T>$T_c$, the normal phase, by \eqref{z metric} and \eqref{normal phase metric 1}, our $g_{yy}$ and $g_{zz}$ will become:
\begin{equation}
    g_{yy}=\frac{1}{z^2},\quad
    g_{zz}=\frac{1}{z^4(\frac{1}{z^2}-\frac{\beta^2}{2}-z(1+\frac{\mu^2}{4}-\frac{\beta^2}{2})+\frac{z^2\mu^2}{4})}
\end{equation}
Substitute them into \eqref{minimal surface 2}, we will get:
\begin{equation}
\label{eq:symmetry eop}
    A=\int_{z_1}^{z_2}\frac{1}{z^2}\sqrt{\frac{1}{1-\frac{z^2\beta^2}{2}-z^3(1+\frac{\mu^2}{4}-\frac{\beta^2}{2})+\frac{z^4\mu^2}{4}}}dz
\end{equation}
Which is the analytic equation of the minimum surface region of the normal phase symmetry configuration cases. We can see this in Fig~\ref{symmetry check test}. We checked it by approaching our configuration to symmetry both T<$T_c$ and T>$T_c$ and calculated the EoP with our method. We also calculated the EoP for symmetry cases by the analytic equation mentioned above, we found out that two method will match when the configuration becomes symmetry. 

\subsubsection{MI-EoP inequality}
\begin{figure}[h!]
    \centering
    \includegraphics[scale=0.6]{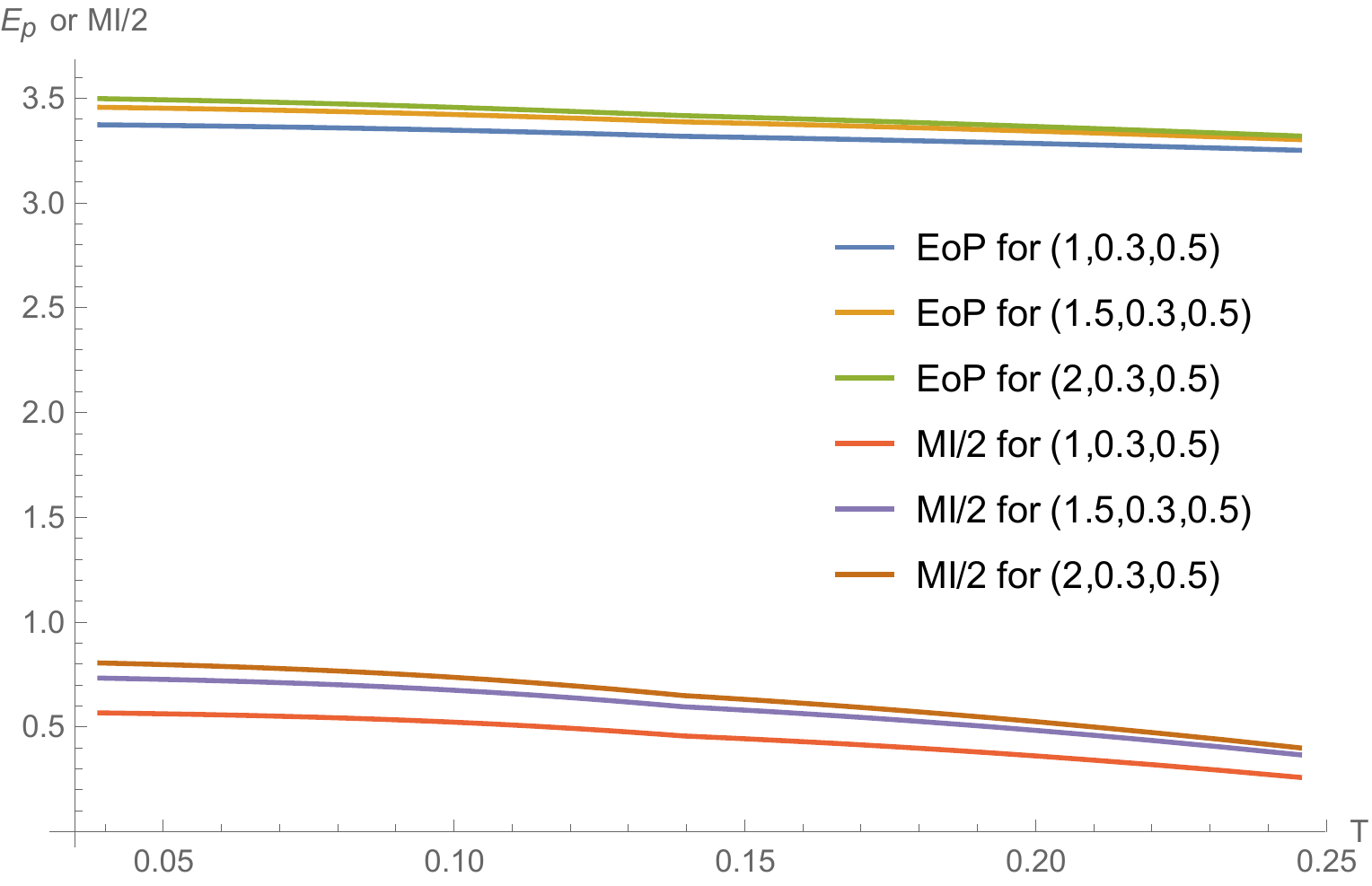}
    \caption{MI/2 are always smaller than EoP, neglect of configuration and temperature. There are kink points at the critical temperature in every line, but the range is too big so they are hard to see.}
    \label{mi check}
\end{figure}

By the inequality of EoP and MI/2 mentioned in Sec~\ref{Holographic entanglement of purification}:
\begin{equation}
\label{eop mi inequality}
    E_p(\rho_{AB})\geq \frac{MI(A:B)}{2}
\end{equation}
which has been proved in \cite{Takayanagi:2017knl,Liu:2019qje}, we know our MI/2 has to be smaller than EoP, it can be easily shown in Fig~\ref{mi check}. MI/2 will always smaller than EoP, neglect of configuration and temperature. 

\subsubsection{High-temperature behavior}
\label{high-temperature behavior}
\begin{figure}[h!]
    \centering
    \includegraphics[scale=0.6]{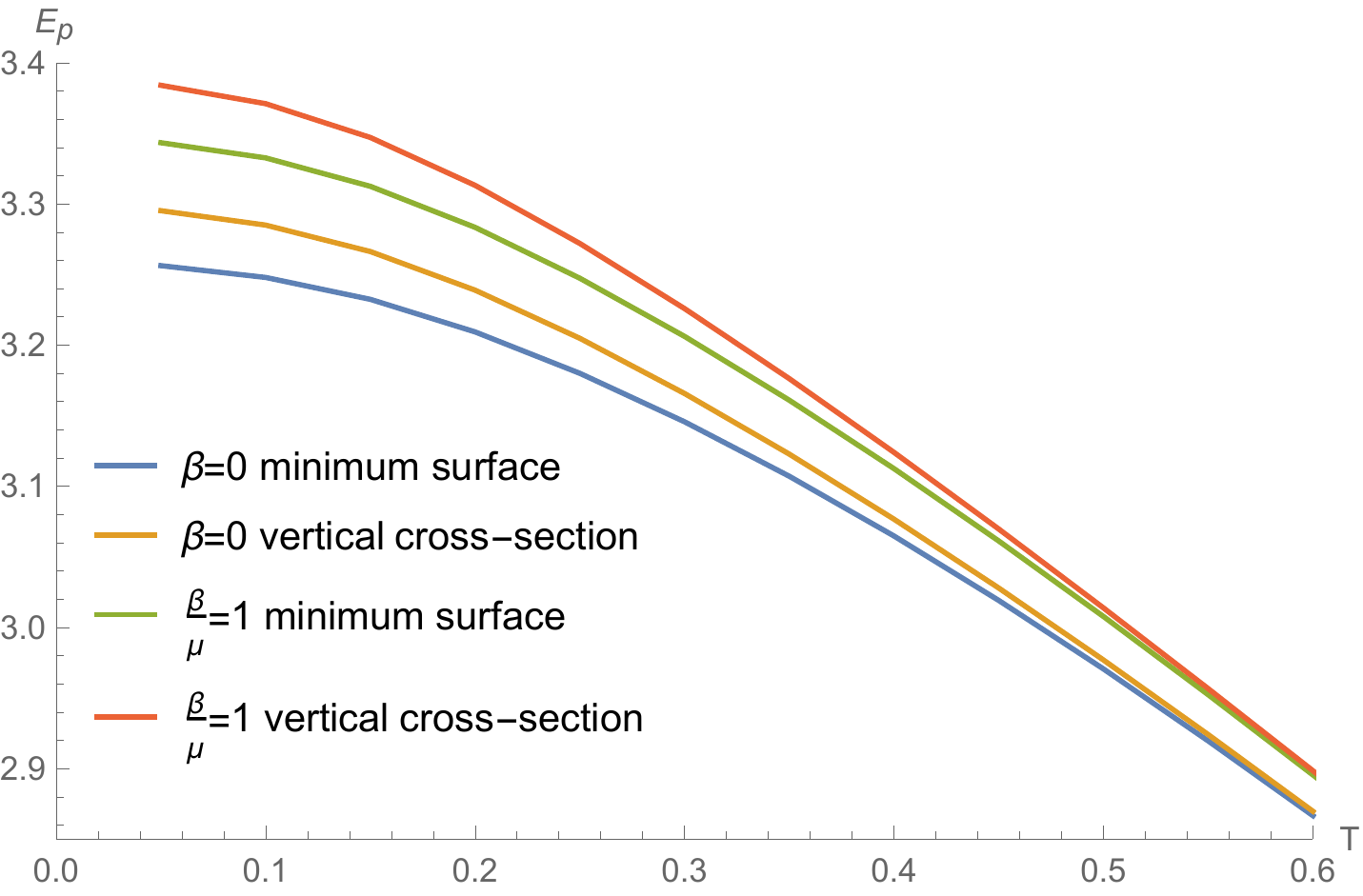}
    \caption{EoP in high-temperature situation for fixed (1,0.3,0.5). The EoP will approach to the straight line minimal cross-section solution in high temperature situation.}
    \label{high-temperature check}
\end{figure}

For high-temperature cases, by the black hole thermodynamic, we know the black hole horizon will get bigger and get closer to the boundary as the temperature going higher. So the distinct between straight line EOM and others oblique EOM will become smaller. After we rescale our $z_h=1$ for all temperature, this is same as pulling the minimum surface for high-temperature very close to horizon. Then the shape of the $C_{abc}$ near the horizon will be very flat, this will cause the minimum cross-section close to straight line. So we claim our EoP behavior will approaches to straight line minimum cross-section behavior in the high-temperature cases. Notice that we can borrow the straight line analytic equation above, although in the non-symmetry cases this will not be the actual minimum surface, but just a good standard to approach. We show our result in Fig~\ref{high-temperature check}.

\subsection{The (a,b,c) configuration}
\label{(a,b,c) configuration}
\begin{figure}[h!]
    \centering
    \includegraphics[scale=0.57]{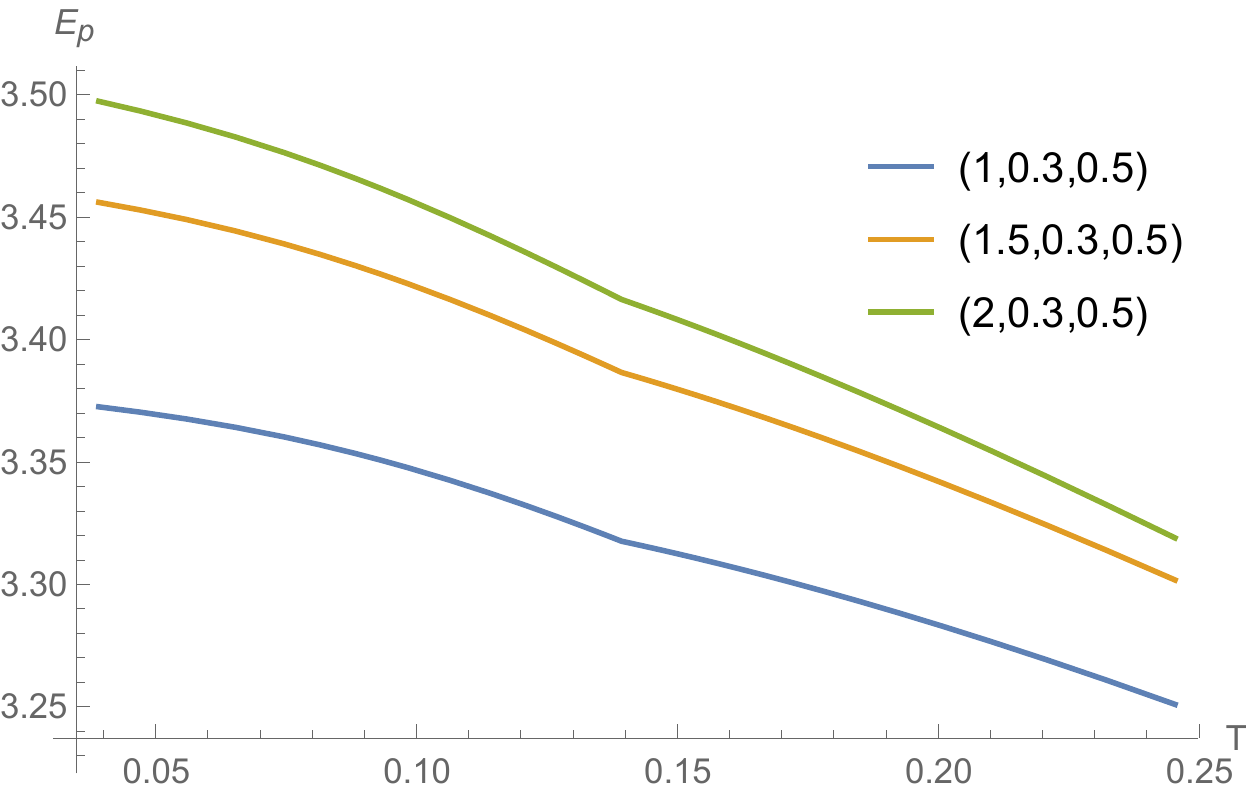}
    \includegraphics[scale=0.57]{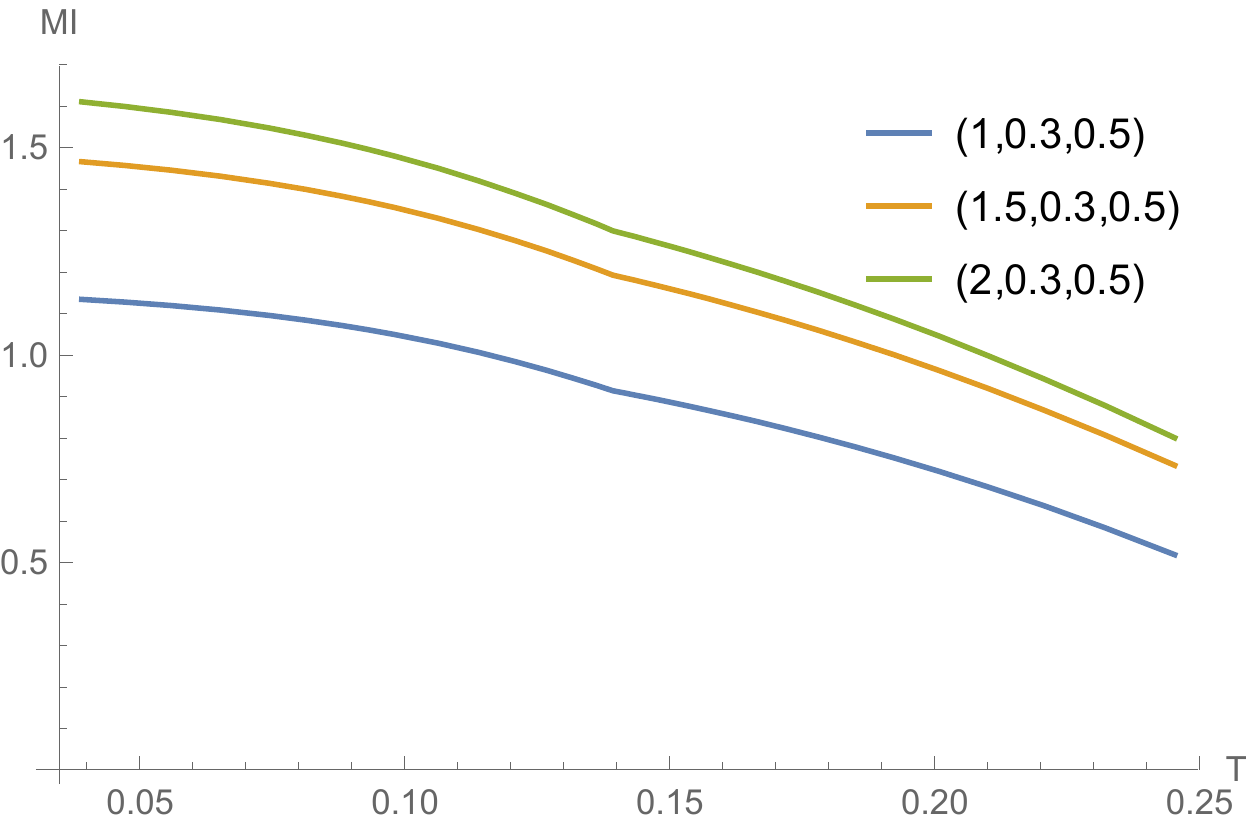}
    \caption{EoP and MI for fixed (a,0.3,0.5) with different "a". Both EoP and MI increase as "a" increase.}
    \label{compare a}
\end{figure}

\begin{figure}[h!]
    \centering
    \includegraphics[scale=0.57]{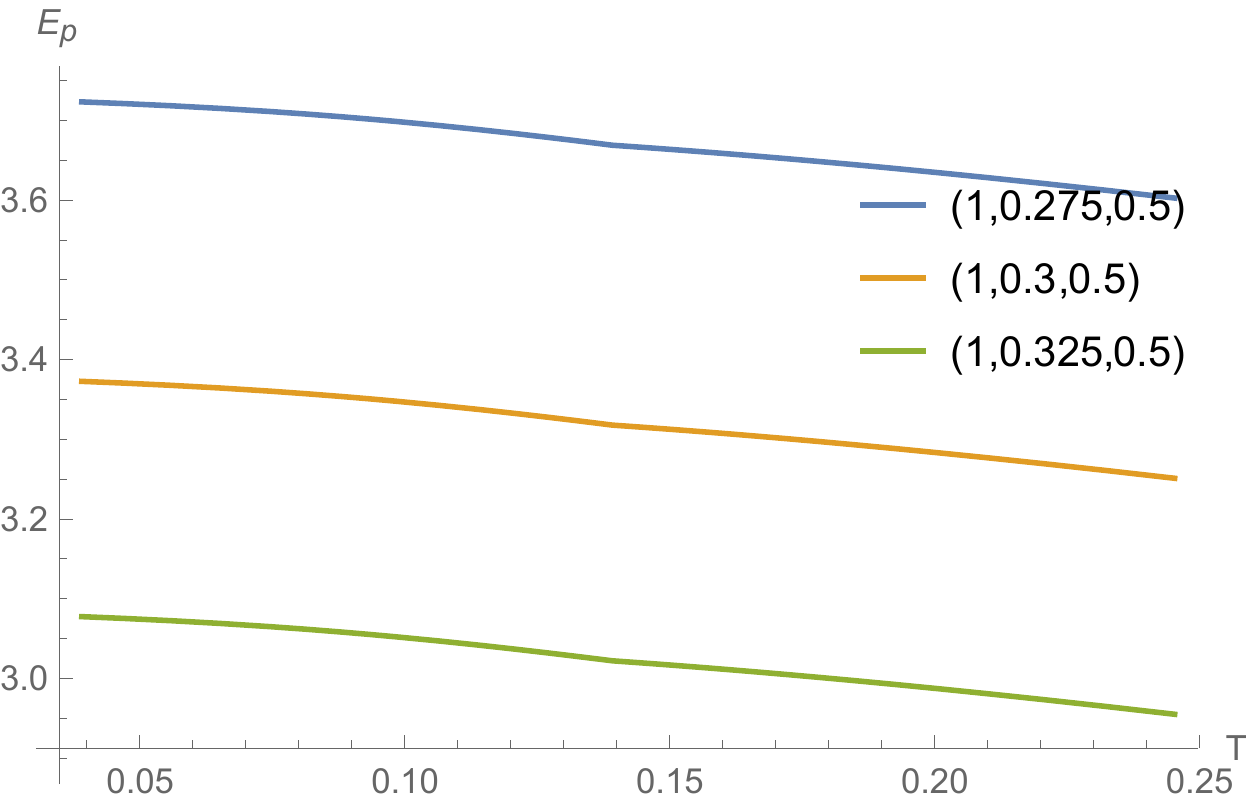}
    \includegraphics[scale=0.57]{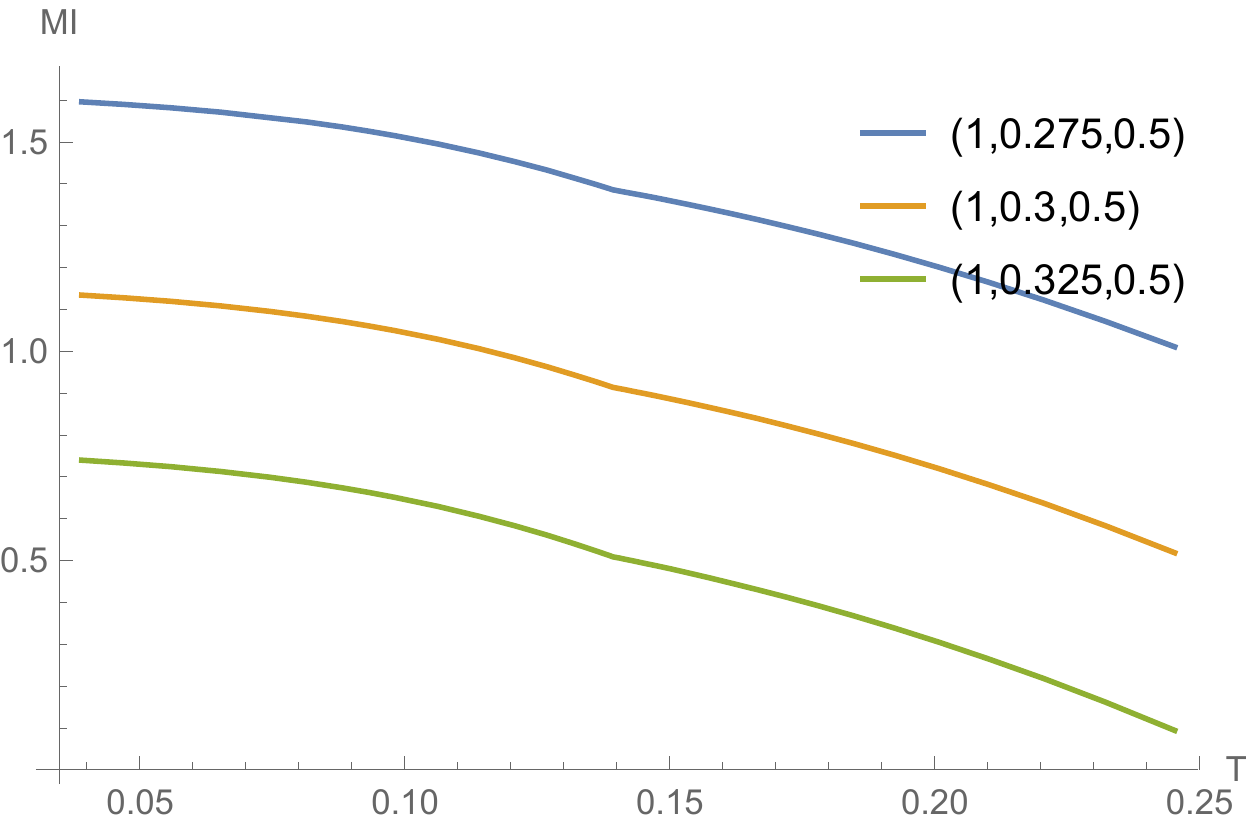}
    \caption{EoP and MI for fixed (1,b,0.5) with different "b". Both EoP and MI decrease as "b" increase.}
    \label{compare b}
\end{figure}

\begin{figure}[h!]
    \centering
    \includegraphics[scale=0.57]{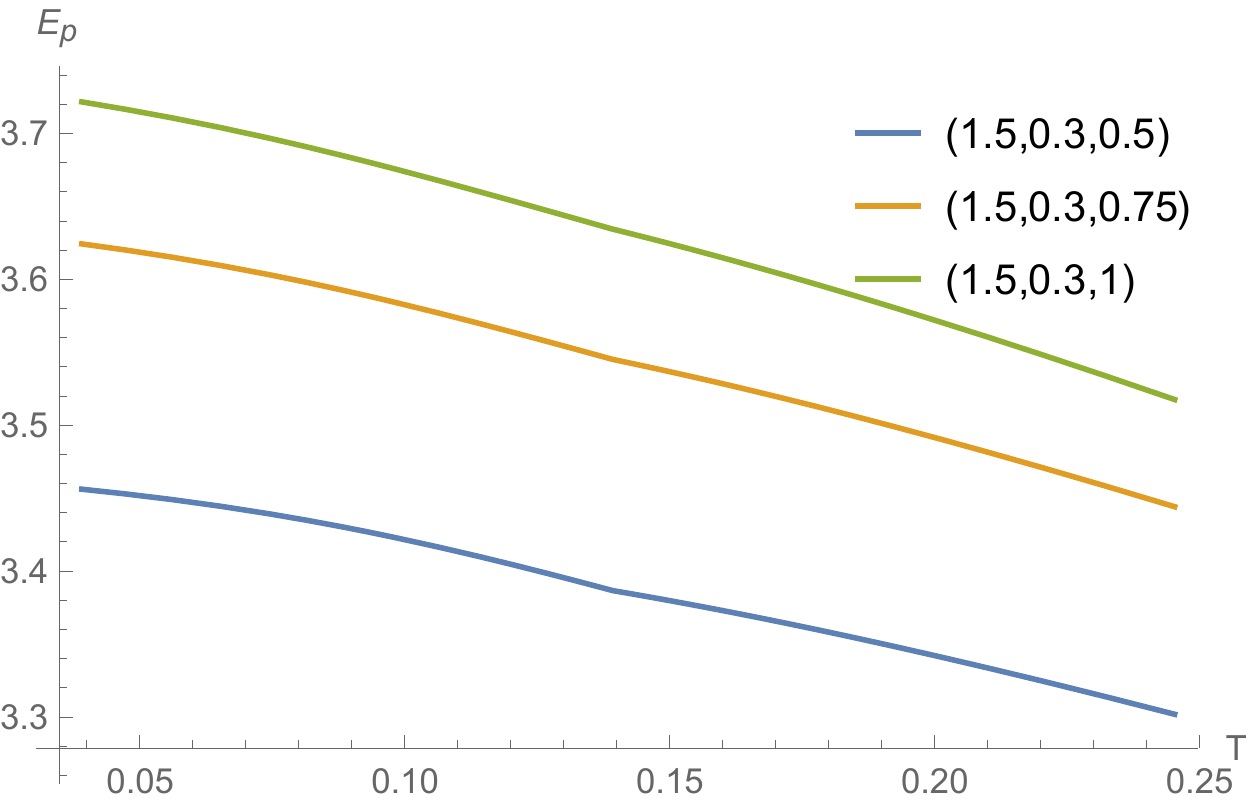}
    \includegraphics[scale=0.57]{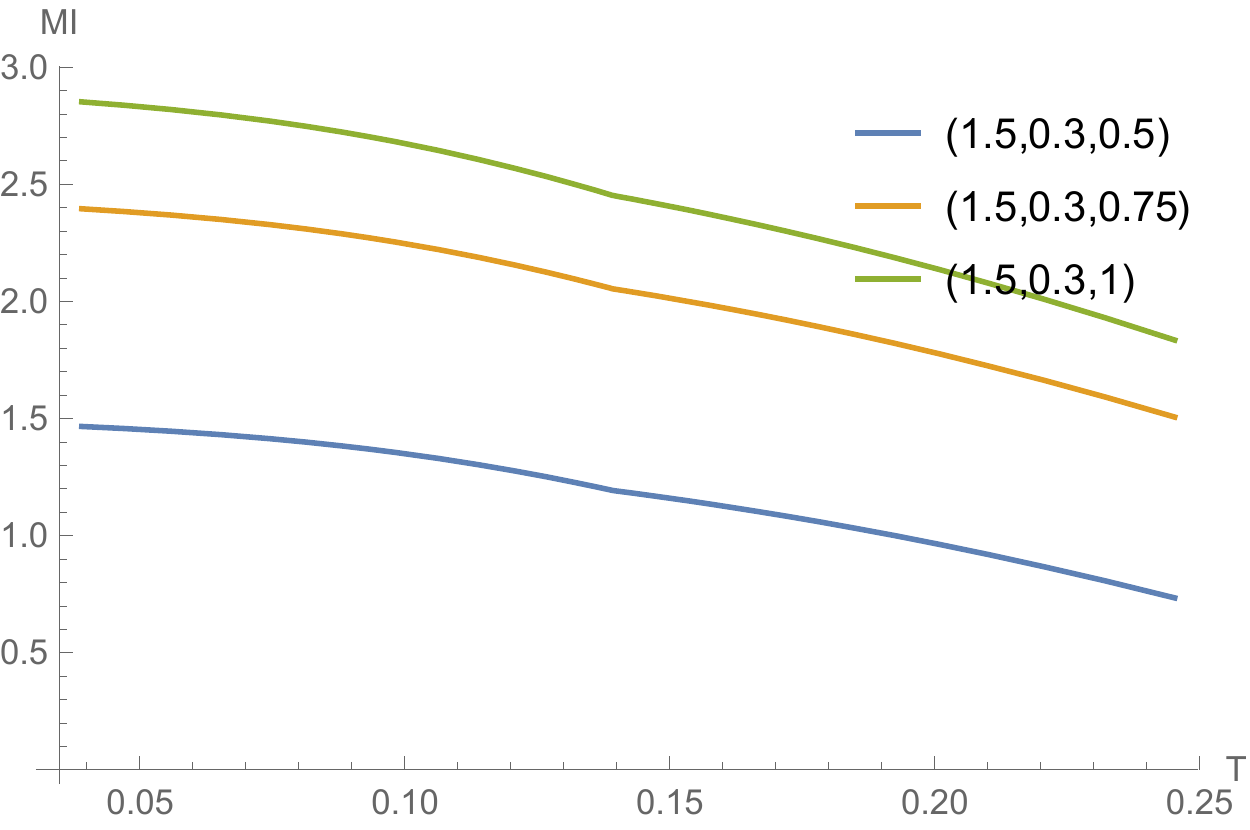}
    \caption{EoP and MI for fixed (1.5,0.3,c) with different "c". Both EoP and MI increase as "c" increase.}
    \label{compare c}
\end{figure}

In this section, we discuss the effect that (a,b,c) configuration can do to EoP. We can see in Fig~\ref{compare a} that by increasing "a", the EoP will increase. Compare to "a", you can see in Fig~\ref{compare b}, the EoP and MI are far more sensitive to the change of "b" than "a". We only increase 0.025 in "b", then the EoP decrease a lot. You can see in Fig~\ref{compare c} that the change of "c" will also not that sensitive than the change of "b". But compare the change of "a" and "c", we notice that since we have restricted the minimal-surface to happen at the "c" side for simplicity, the influence of "a" is much less than the one of "c", which is quit obvious in the geometrical view, or Fig~\ref{fig:entanglement wedge}. You can also notice that the dependence on the configuration doesn't change a lot in the normal phase and superconductor phase. It is hard to diagnose the critical temperature by the configuration dependence.

There is another inequality for EOP we can check in this part of discussion, mentioned in \cite{Takayanagi:2017knl,Liu:2019qje}, and have been geometrically proved in \cite{Liu:2019qje}:
\begin{equation}
    E_p(a,b,c+\delta c)\geq E_p(a,b,c), \quad \delta c\geq 0.
\end{equation}
which can be easily checked in the left plot of Fig~\ref{compare c}. As "c" become larger, the EoP will always increase.

We also discuss the behavior of MI due to the (a,b,c) configuration, which are showed in the right plots of Fig~\ref{compare a},\ref{compare b} and \ref{compare c}. We can find out that they are similar to EoP. We conclude these results that increasing the subsystem's region will let the entanglement of two subsystem increase, but increasing their distance will let the entanglement decrease.

\subsection{Momentum Relaxation Effect}
\label{beta Effect}
\subsubsection{EoP and MI}

\begin{figure}[h!]
    \centering
    \includegraphics[scale=0.57]{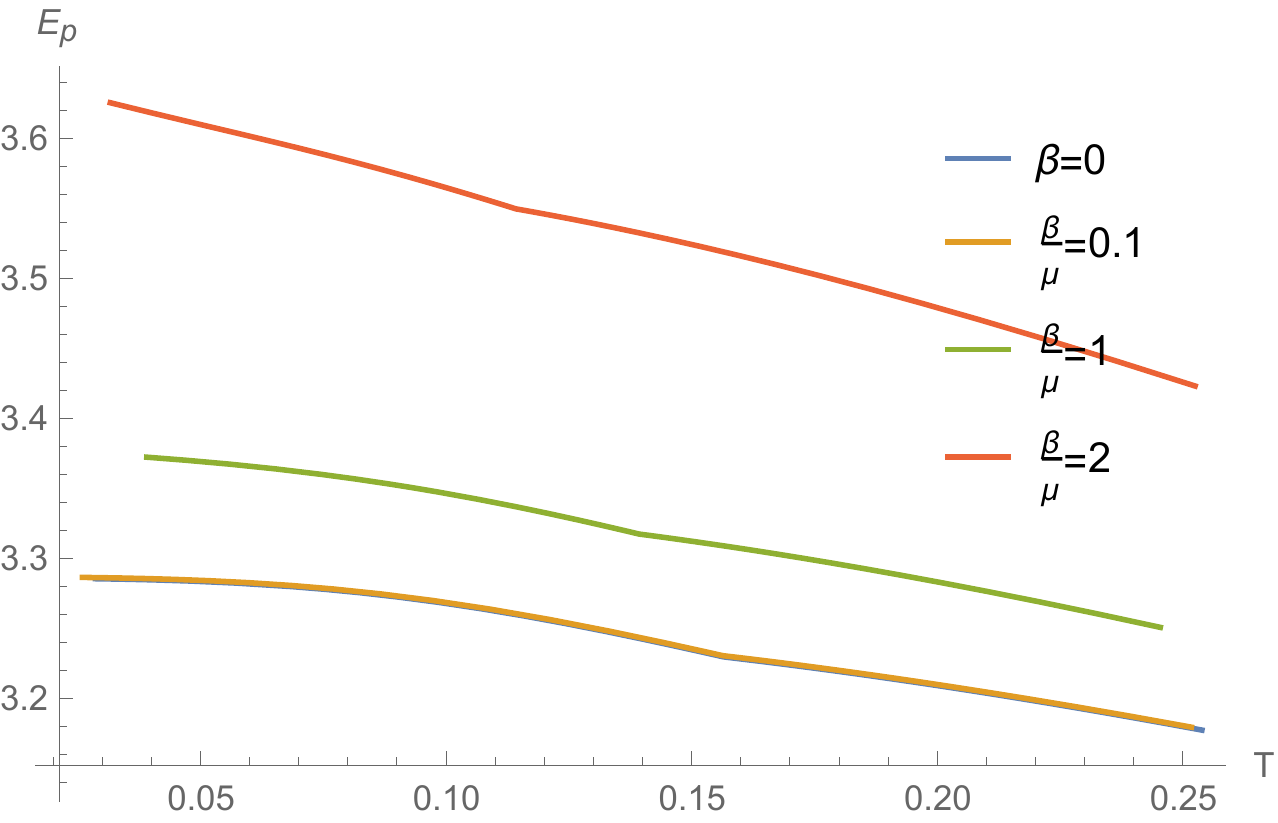}
    \includegraphics[scale=0.57]{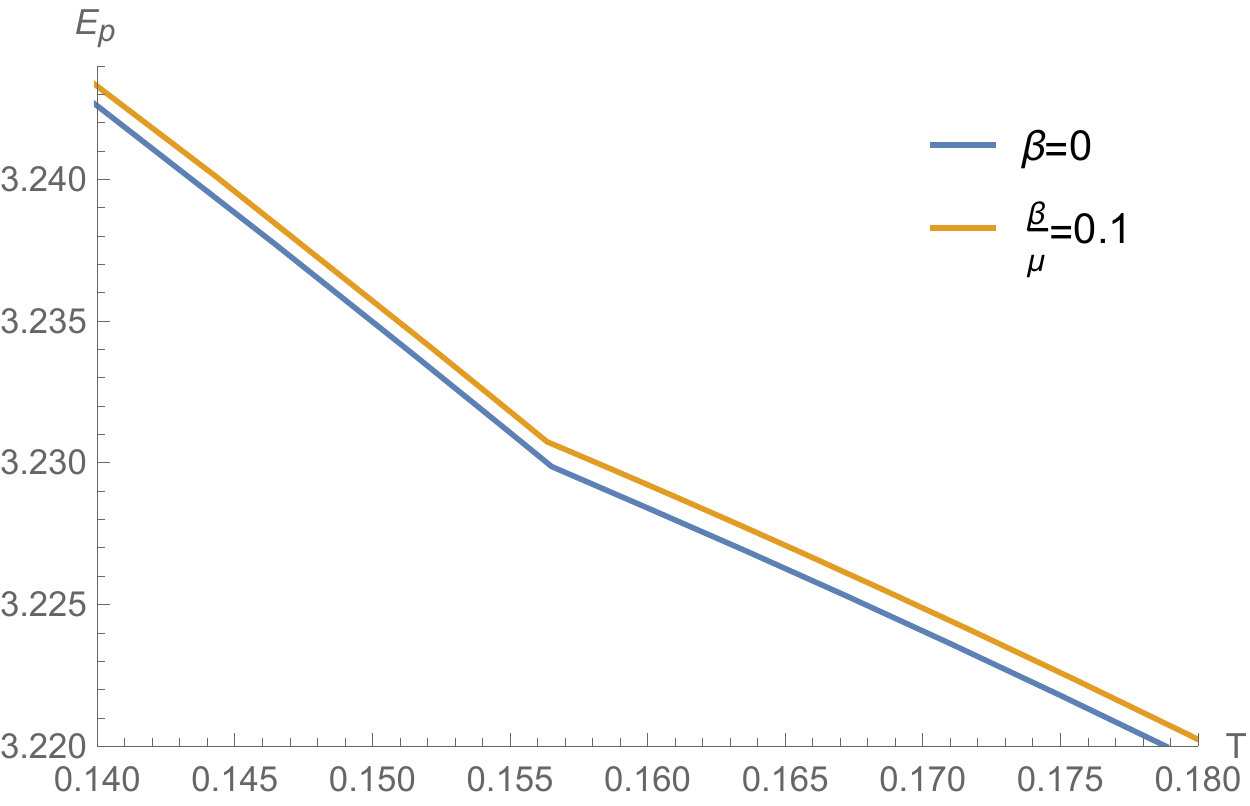}
    \caption{EoP for fixed (1,0.3,0.5) with different $\beta/\mu$. In the right plots we scale out the $\beta=0$ and $\beta/\mu=0.1$ cases. As the momentum relaxation effect become stronger, the EoP will become bigger.}
    \label{compare mu b}
\end{figure}

\begin{figure}[h!]
    \centering
    \includegraphics[scale=0.57]{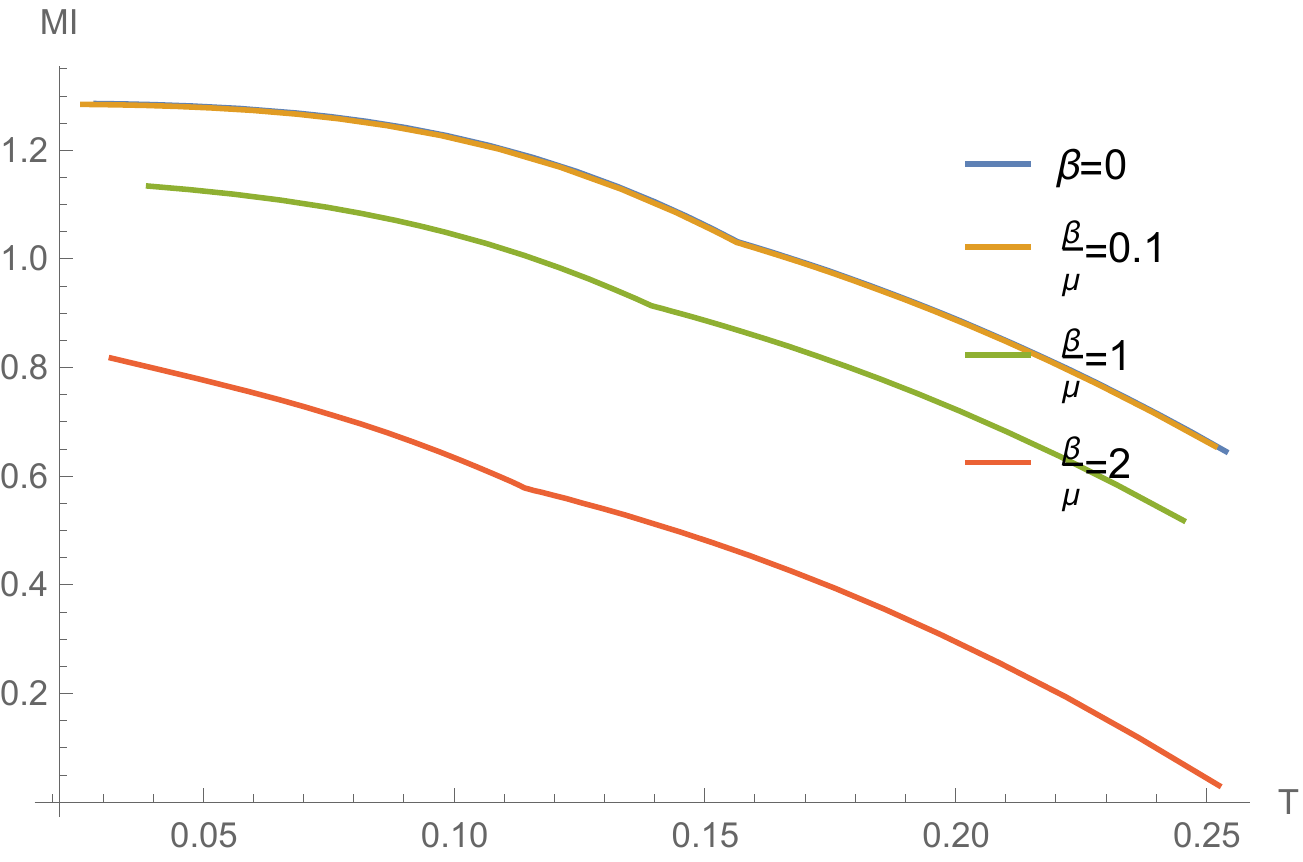}
    \includegraphics[scale=0.57]{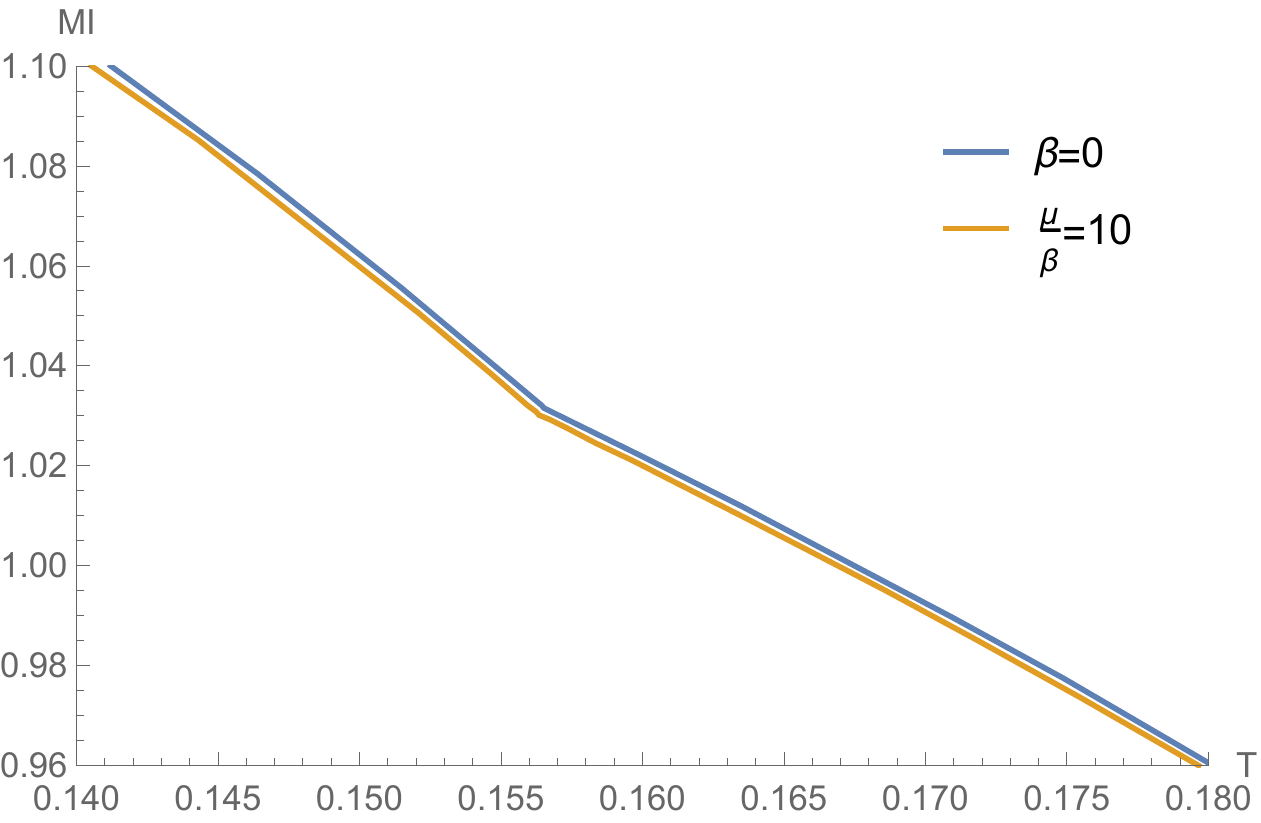}
    \caption{MI for fixed (1,0.3,0.5) with different $\beta/\mu$. In the right plots we scale out the $\beta=0$ and $\beta/\mu=0.1$ cases. As the momentum relaxation effect become stronger, the MI will become smaller.}
    \label{mi compare mu b}
\end{figure}

\qquad Finally, we focus on how the momentum relaxation affect the EoP and MI. By Fig~\ref{compare mu b}, we see that with $\beta/\mu$ increase, which means the momentum relaxation becomes stronger, the EoP will increase. But shown in Fig~\ref{mi compare mu b}, we can see that the MI will decrease with the momentum relaxation become stronger. This phenomenon is both established in normal phase and superconductor phase. In normal phase, this is actually very similar to the case in \cite{Huang:2019zph}, which also claimed that momentum relaxation will affect this two quantum information quantities opposite in some configuration. But in our case, we figure out that in superconductor phase this phenomenon will last. This is quit interesting, unlike the relation they have in (a,b,c) configuration, which they behave in the same way. This means the EoP and MI content different information. Keep in mind that this opposite behavior also guarantee the inequality \eqref{eop mi inequality} will not be broke as the momentum relaxation become stronger.

\begin{figure}[h!]
    \centering
    \includegraphics[scale=0.57]{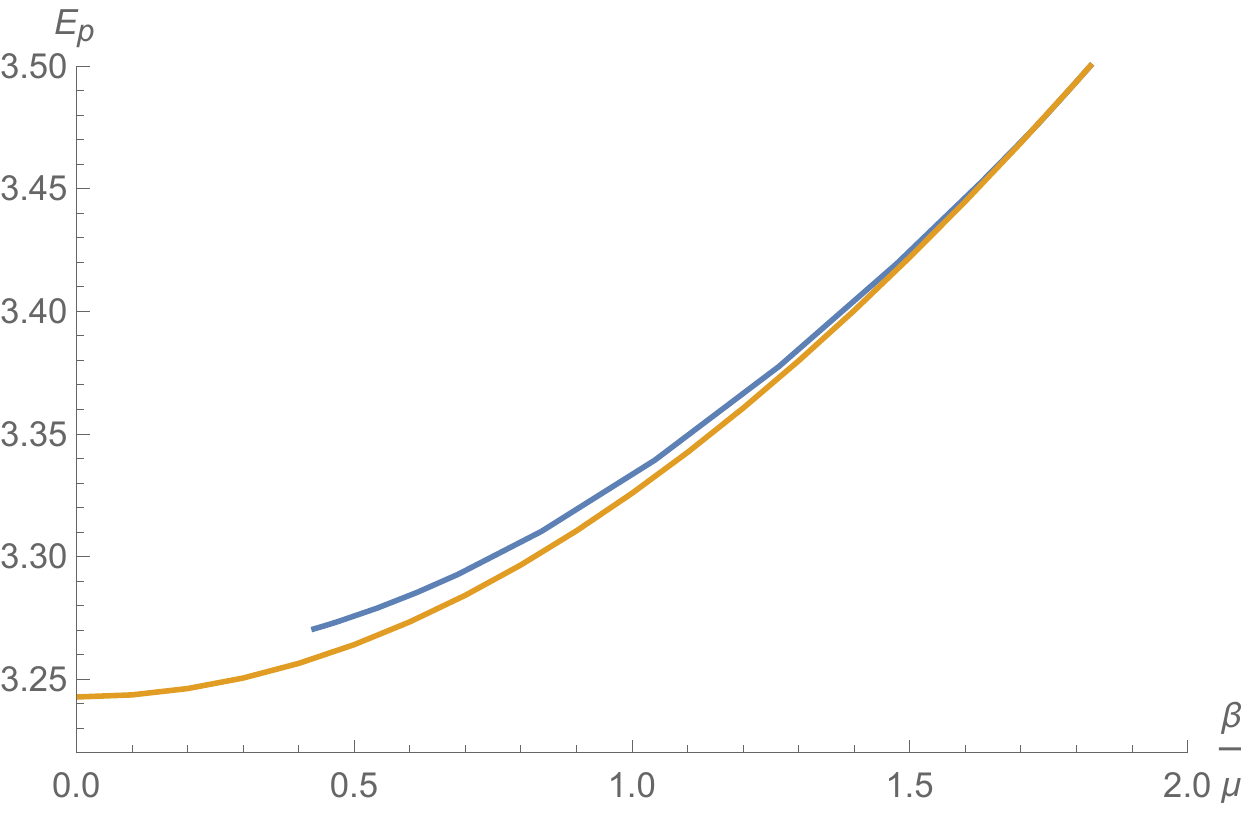}
    \includegraphics[scale=0.57]{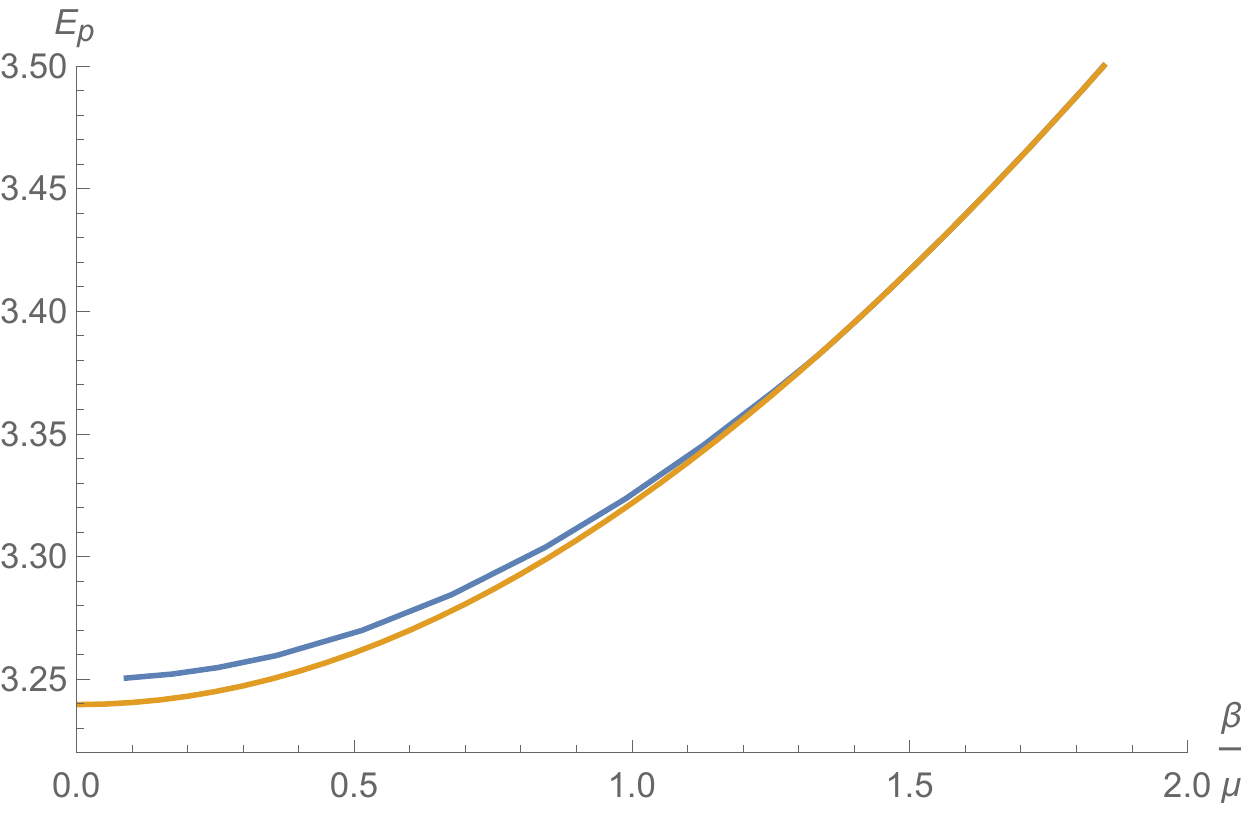}
    \includegraphics[scale=0.57]{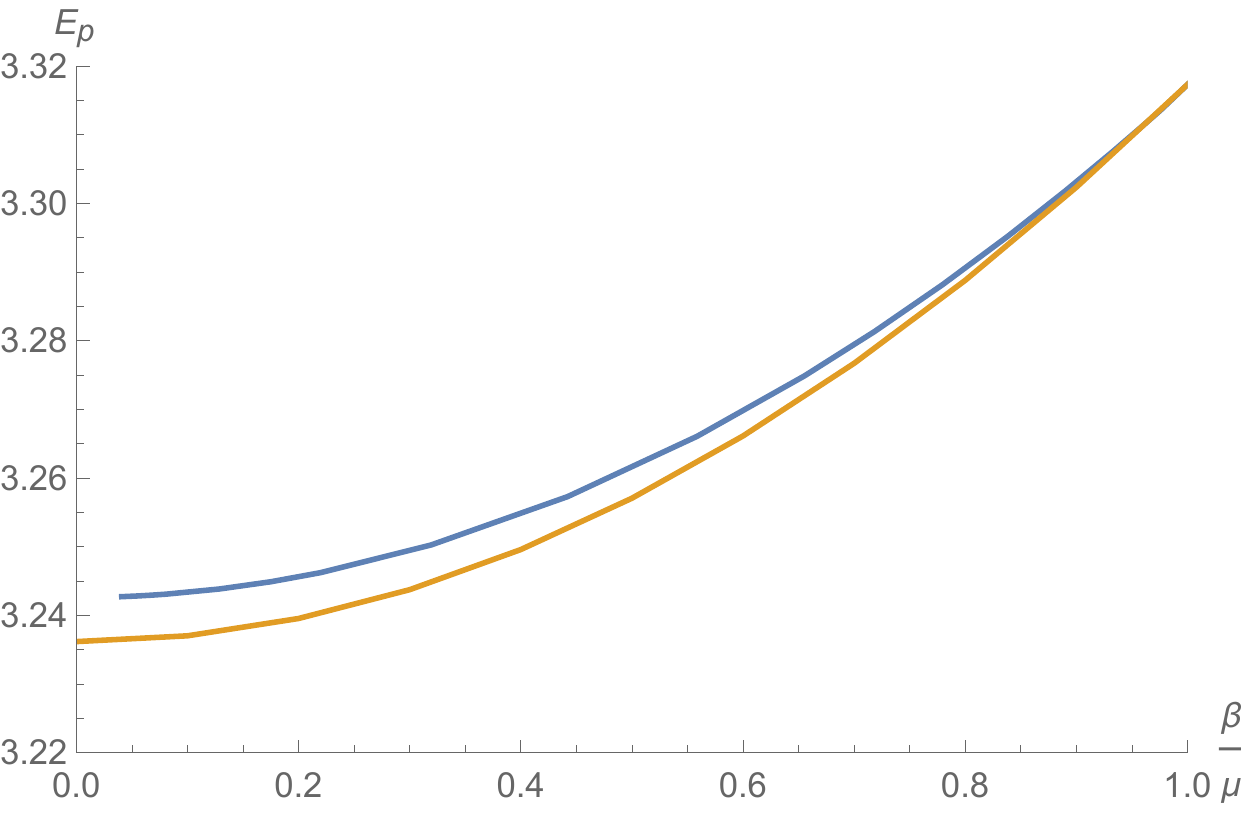}
    \includegraphics[scale=0.57]{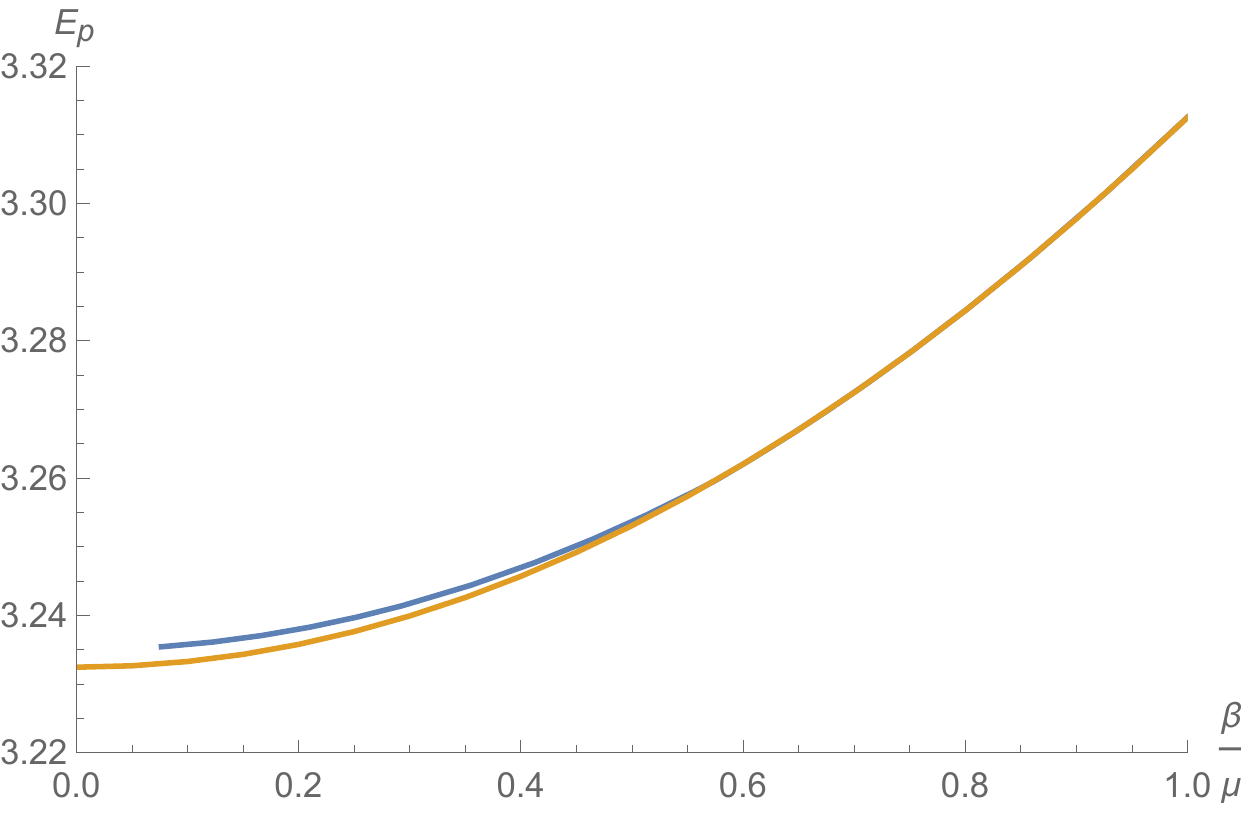}    
    \caption{EoP for fixed (1,0.3,0.5) with different T. From upper left to lower right are T=0.12,0.13,0.14,0.15.}
    \label{T fixed compare b mu}
\end{figure}

The superconductor phase, however, is still different than the normal phase. It will change the behavior of the EoP and MI effected by momentum relaxation. We can see this phenomenon in two difference ways. First, in Fig~\ref{T fixed compare b mu}, the EoP and MI will be larger than the case without the superconductor, which means the superconductor effect is actually enhance the entanglement degree, and actually, this can also be observed in the discussion of (a,b,c) configuration, the superconductor phase will always let both EoP and MI bigger than the case without superconductor. In the second way, the superconductor effect is actually slightly suppress the effect that the momentum relaxation do to EoP, because the distinct between ($\frac{\beta}{\mu}=0\backsim\frac{\beta}{\mu}=2$) is smaller now, but on the other hand, slightly enhance the effect do to MI, since the distinct of MI between this region is increase.

\begin{figure}[h!]
    \centering
    \includegraphics[scale=0.57]{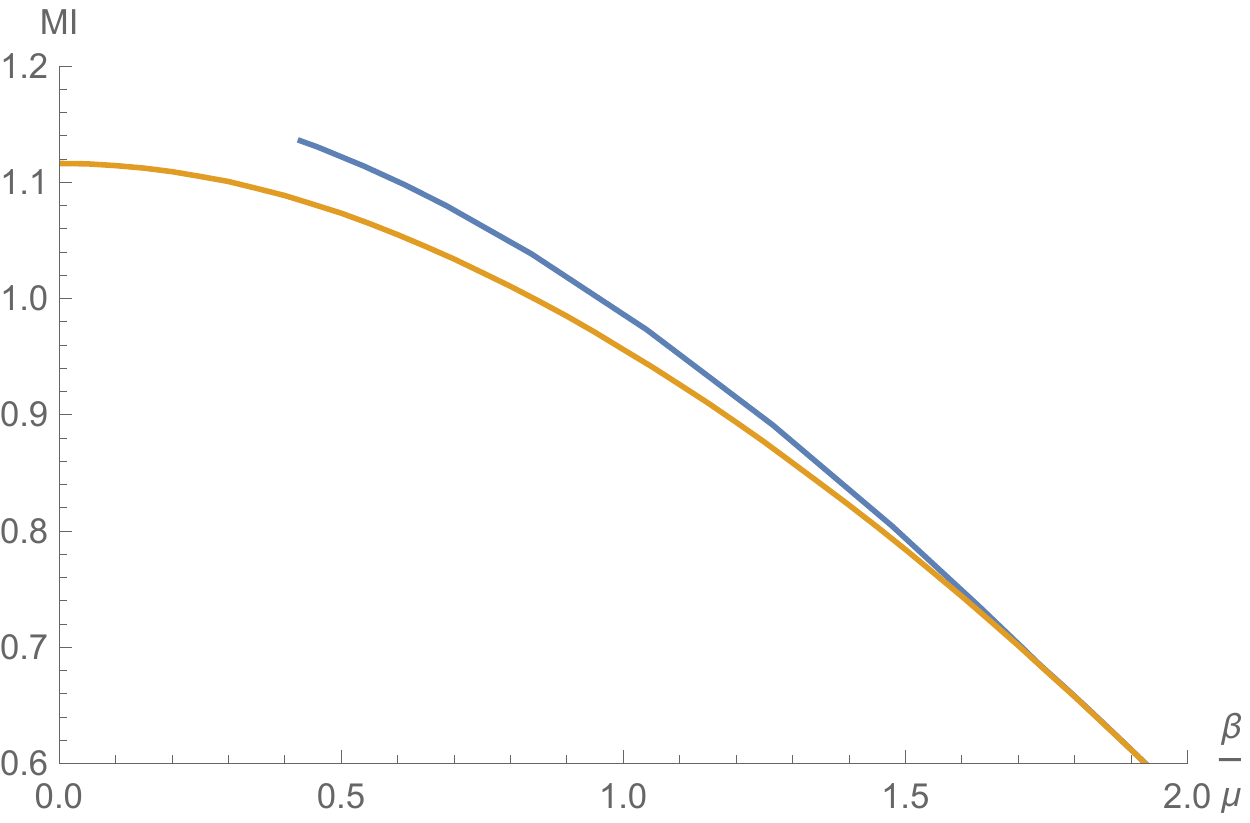}
    \includegraphics[scale=0.57]{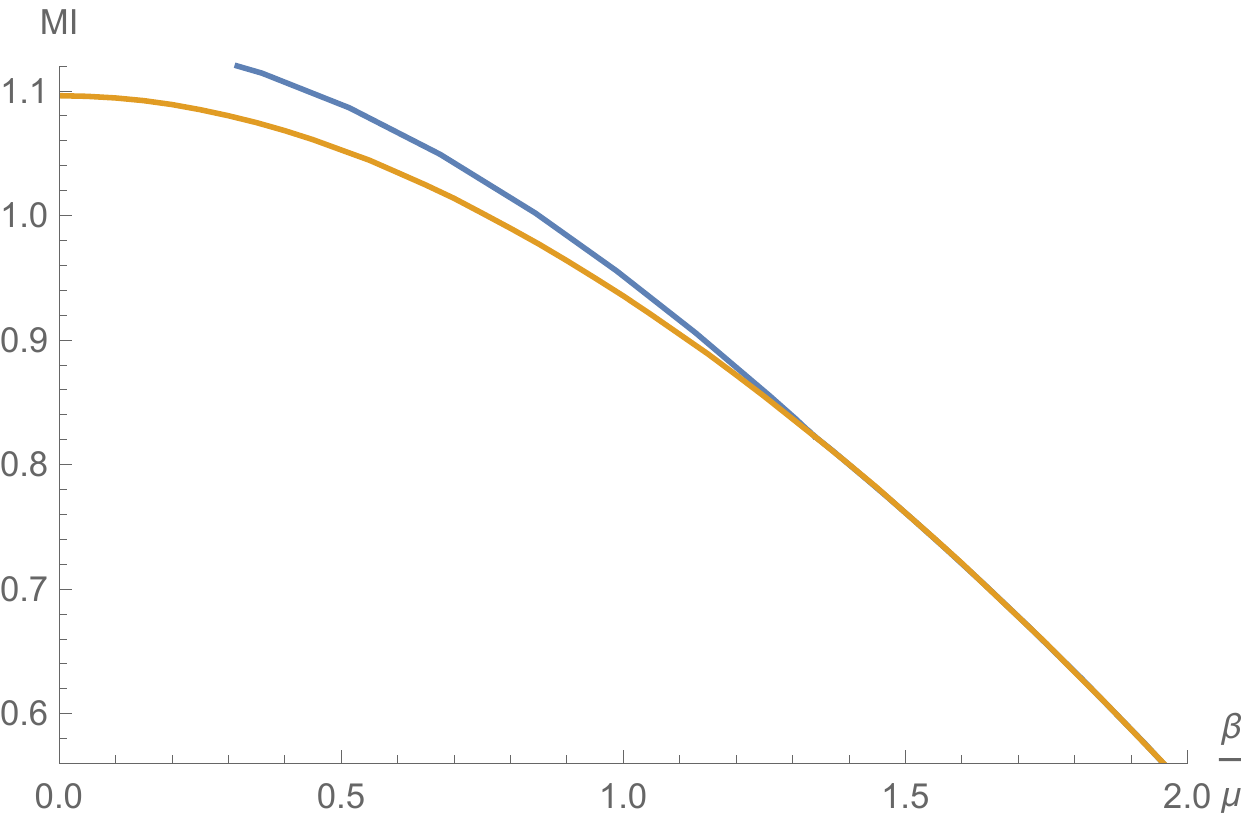}
    \includegraphics[scale=0.57]{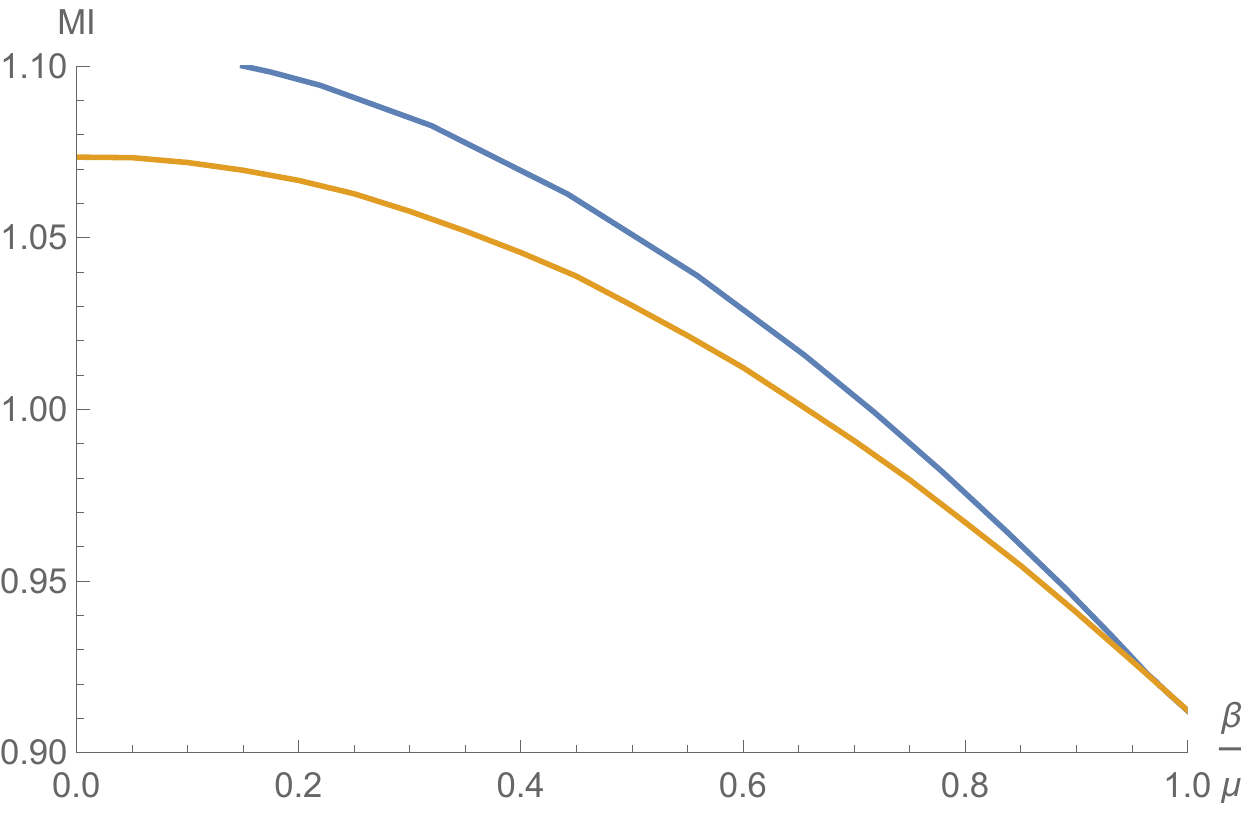}
    \includegraphics[scale=0.57]{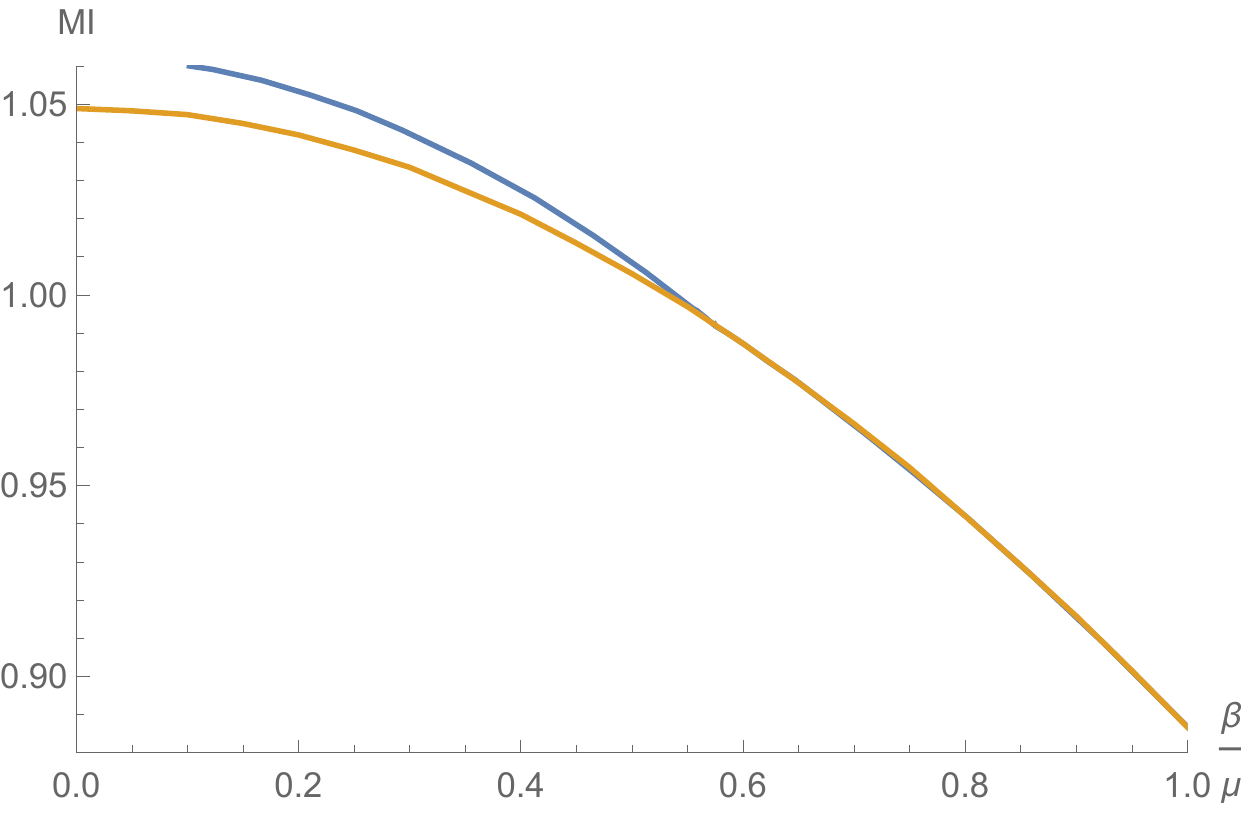}    
    \caption{MI for fixed (1,0.3,0.5) with different T. From upper left to lower right are T=0.12,0.13,0.14,0.15.}
    \label{mi T fixed compare b mu}
\end{figure}

\subsubsection{Critical "c"}
\begin{figure}[h!]
    \centering
    \includegraphics[scale=0.58]{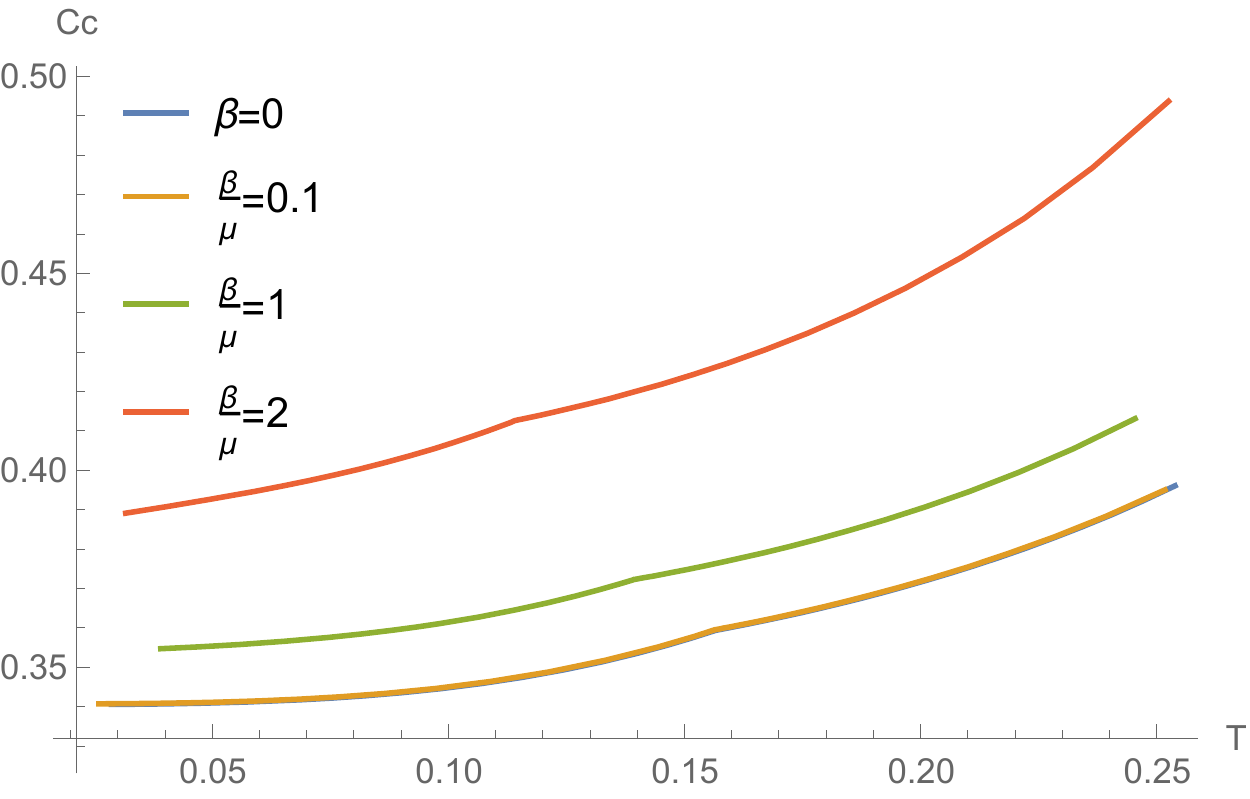}
    \includegraphics[scale=0.57]{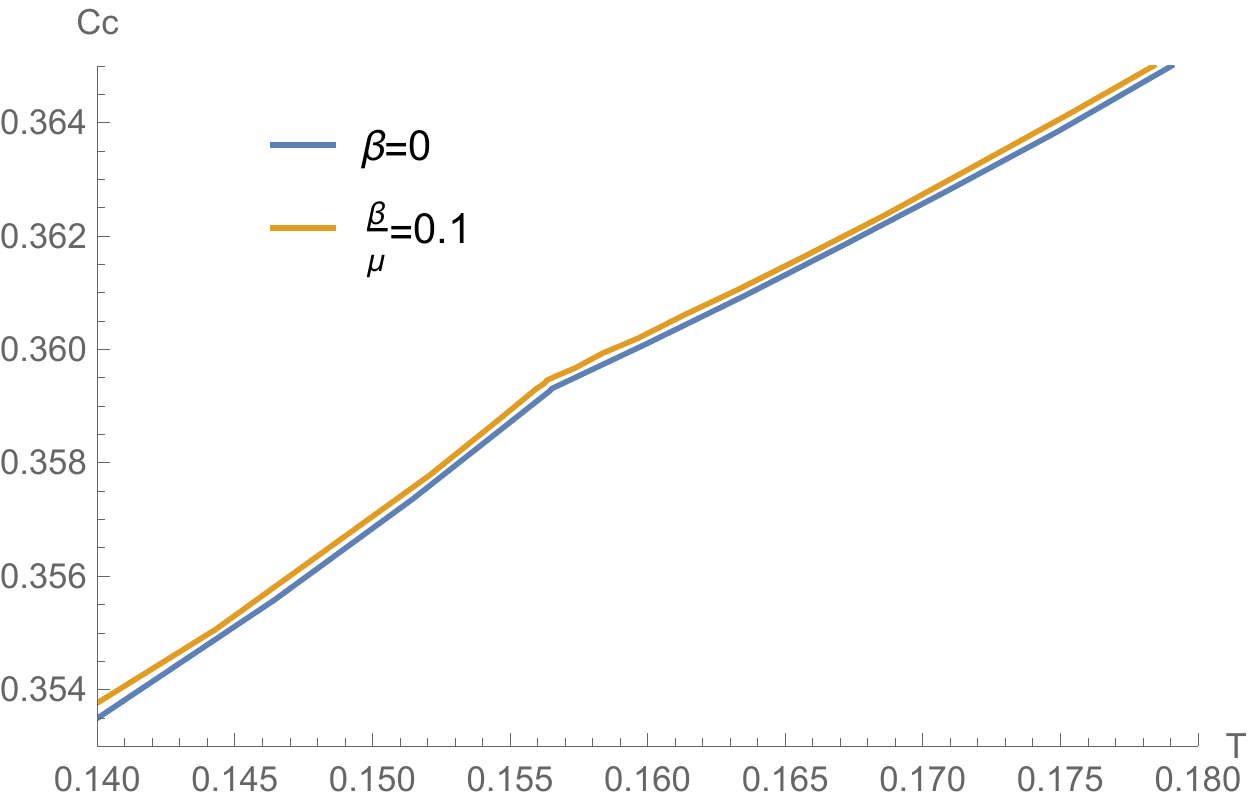}
    \caption{Critical c with fixed (1,0.3,c) in different $\beta/\mu$ situation. In the right plot we scale out the $\beta=0$ and $\beta/\mu=0.1$ cases. As the momentum relaxation effect become stronger, the critical c will become smaller.}
    \label{critical c compare to mu b}
\end{figure}

We can also discuss how the momentum relaxation affect the critical "c", what we mean by critical "c" or just $C_c$ is the critical length of the "c" configuration. Followed \cite{Ryu:2006bv}, we know that MI will not always be positive for all configuration of all temperature, you can see this in the left plot in Fig~\ref{mi compare mu b}, MI is about to vanish in the $\beta/\mu=2$ situation, so we can discuss below which length of "c" with fixed "a","b" will let the MI be zero. It turns out that with the momentum relaxation effect getting stronger, the $C_c$ will increase, which we show in Fig~\ref{critical c compare to mu b}.

\subsubsection{Discussion}
\begin{figure}[h!]
    \centering
    \includegraphics[scale=0.58]{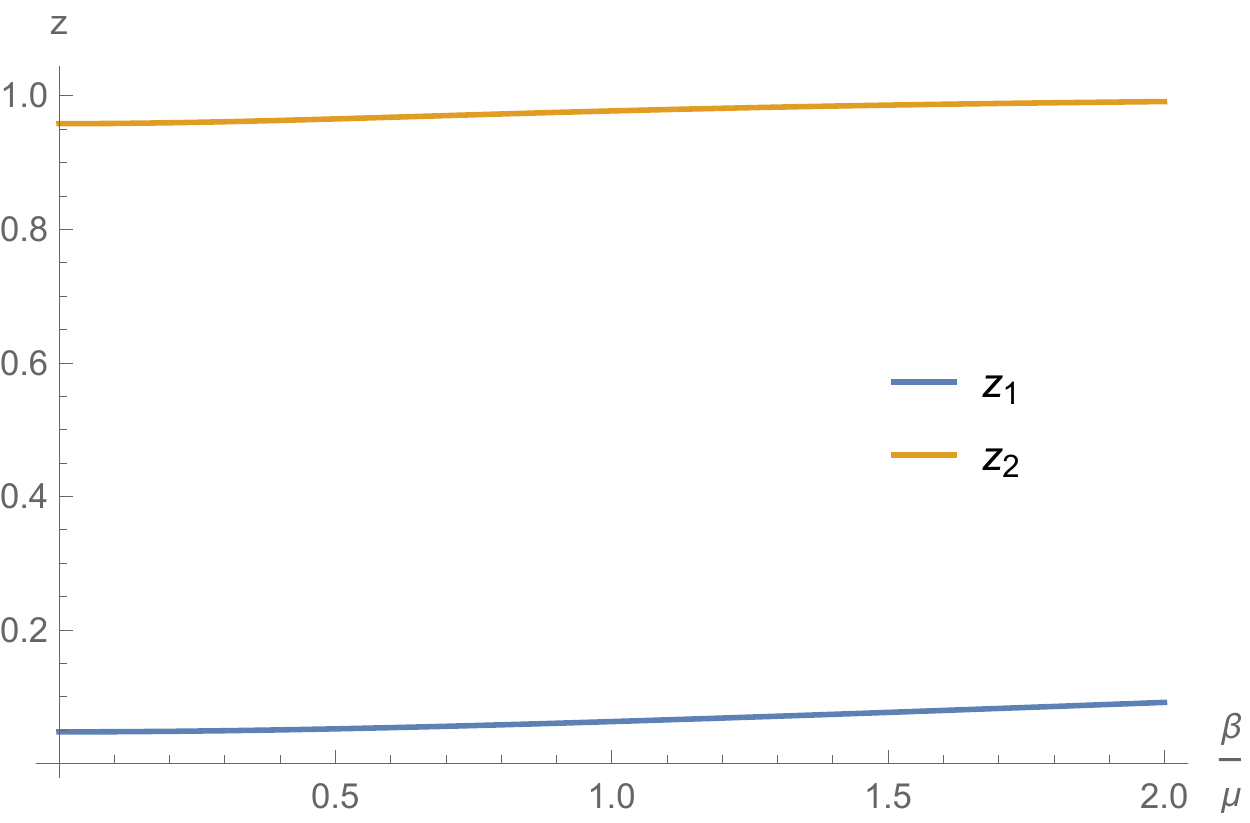}
    \includegraphics[scale=0.58]{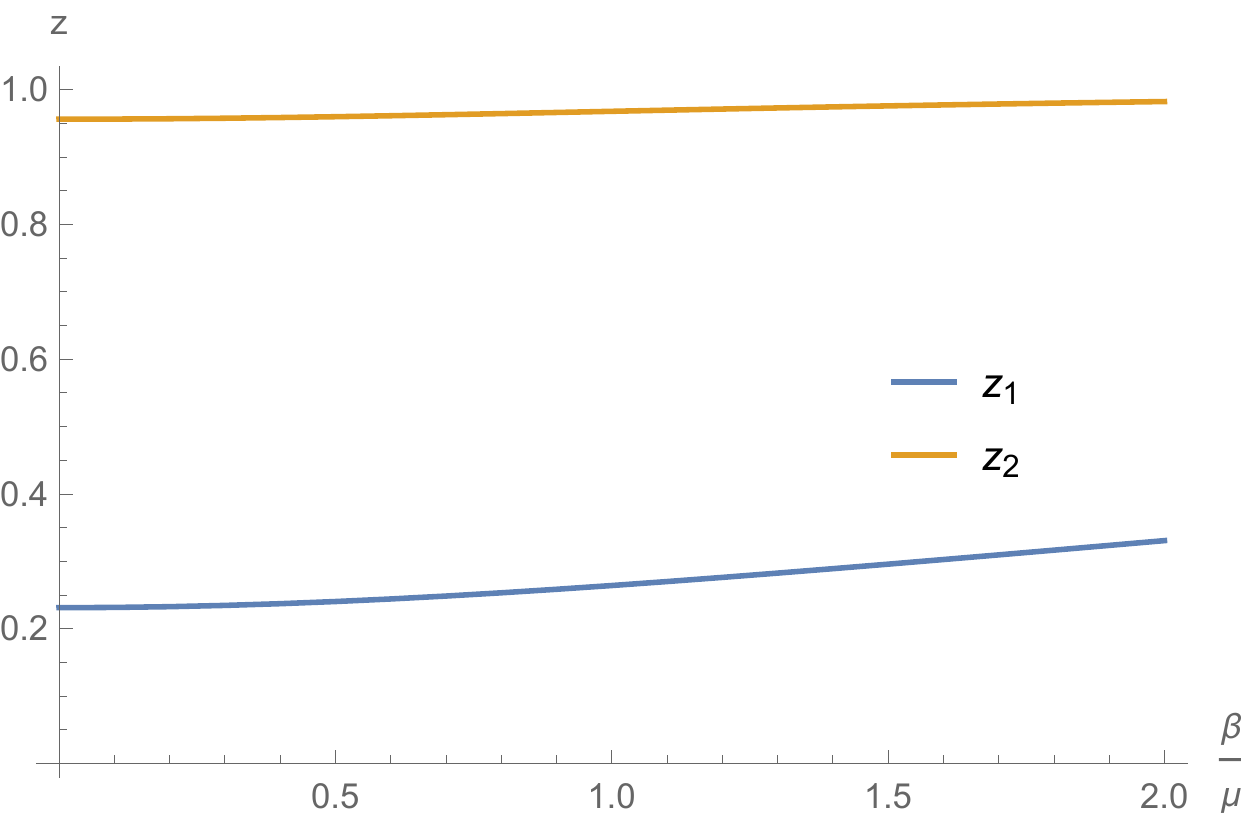}
    \caption{Cross-section length with different configuration and temperature. The left plot is at T=0.1, (a,b,c)=(2,0.1,2) and the right plot is at T=0.2, (a,b,c)=(1,0.3,1). There is no pattern we can tell for cross-section as momentum relaxation increase.}
    \label{fig:cross-section}
\end{figure}

For superconductor phase, it is hard to analyze because all the metric are numerical, instead, it is easier for us to discuss the normal phase since we see the similar behavior on momentum relaxation dependence. For EoP, to simplify, we also take the symmetry case, which is the situation of \eqref{eq:symmetry eop}. Since the shape of the minimum surface will change with the momentum relaxation effect, we can't directly differentiate this equation, instead, we split this into two parts to discuss. First part is the "cross-section length", it is how the "$z_1$" and "$z_2$" distance change with the "$\frac{\beta}{\mu}$", we find that it's hard to see the pattern in this, the length will depends on the temperature and the configuration. We plot different kinds of result in Fig~\ref{fig:cross-section}.
\if
we figure out this distance will decrease as the momentum relaxation becomes stronger. The minimum surface near the horizon will not change much, but the minimum surface near the boundary will change relatively obvious. We can see this in Fig~\ref{fig:cross-section}. This phenomenon is actually consistent with the \ref{high-temperature behavior} that fixed the $z_h=1$ is actually pulling the minimum surface to the horizon. But in this case, we are actually increasing the "$\frac{\beta}{\mu}$" instead of temperature.
\fi

Here we temporarily turn to the discussion of dimensional units, since the fixed $z_h=1$ is dimensional and we want to align the horizons for different $\beta/\mu$ in our plots.

\begin{figure}[h!]
    \centering
    \includegraphics[scale=0.58]{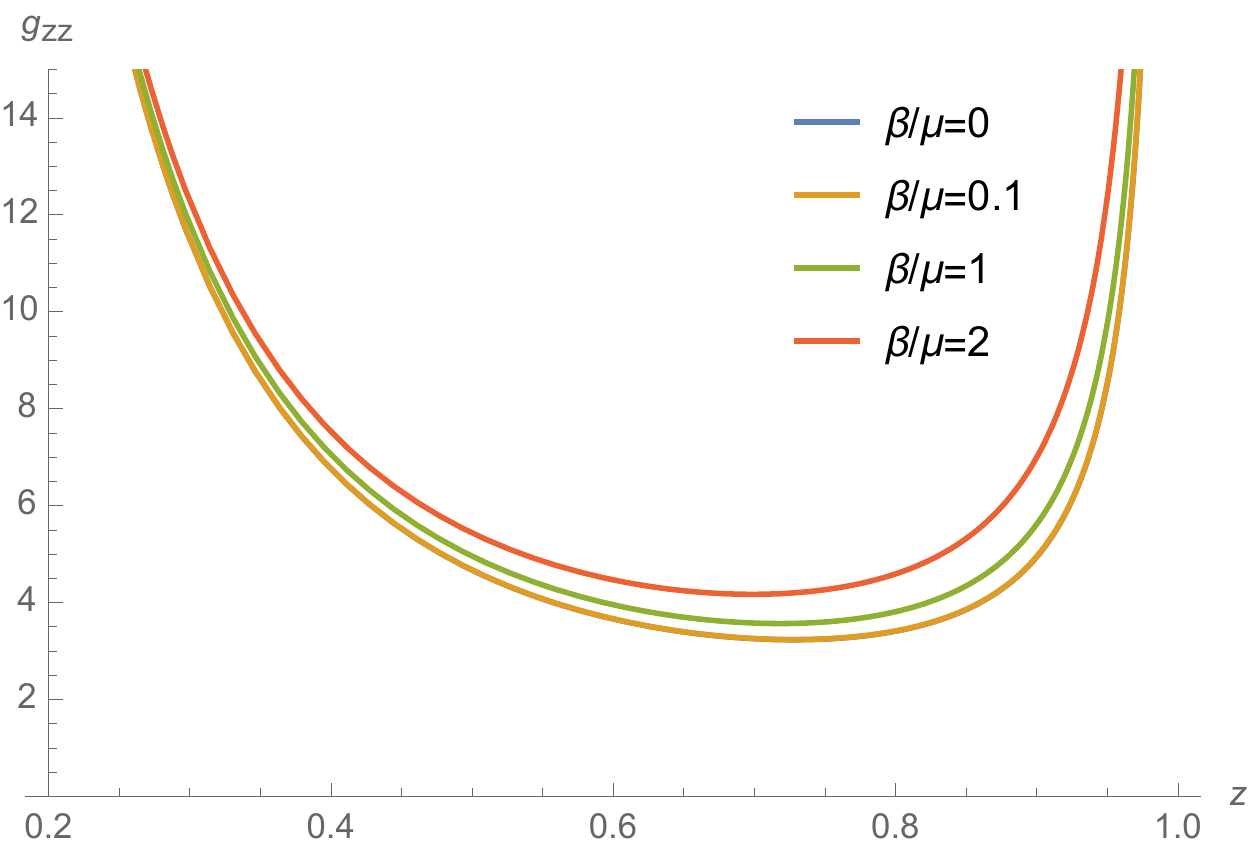}
    \includegraphics[scale=0.58]{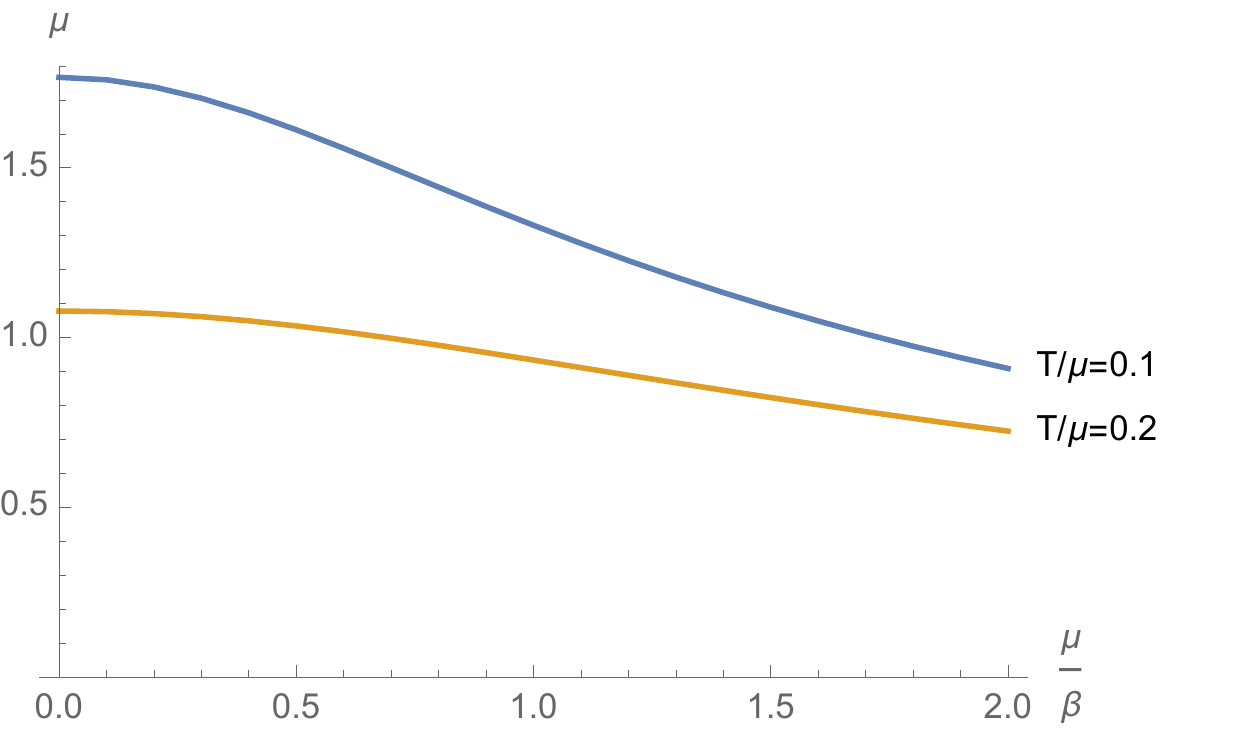}
    \caption{(Left) Metric component $g_{zz}$ at T=0.2. As the momentum relaxation becomes stronger, $g_{zz}$ will become bigger. (Right) The soved $\mu$ decreases as the momentum relaxation becomes stronger.}
    \label{fig:gzz}
\end{figure}

Second part is how the background change with the momentum relaxation. Here we recorded the solved metric and $\mu$, which are independent of $(a,b,c)$ configuration. Since in \eqref{eq:symmetry eop}, the main part dominate the quantity is actually $g_{zz}$ ($g_{yy}$ always stays the same), you can see \eqref{minimal surface 2} more clearly. We plot the result in Fig~\ref{fig:gzz}. It turns out that $g_{zz}$ will increase as the momentum relaxation become stronger.

We find out although the cross-section doesn't has the pattern as the momentum relaxation going stronger, the metric always increase. In addition, the decreasing $\mu$ also let the dimensionless EoP tend to increase. Combine these two parts, the EoP always increase with momentum relaxation seems reasonable.

For MI, since it is the linear combination of HEE, we first discuss the HEE's behavior. Put \eqref{normal phase metric 1} into \eqref{general minimum surface x}:
\begin{equation}
    A=\int_0^w\sqrt{\frac{\frac{4 z'(x)^2}{(z(x)-1) \left(z(x) \left(z(x) \left(2 \beta ^2+\mu ^2
    z(x)-4\right)-4\right)-4\right)}+1}{z(x)^4}}dx
\end{equation}
We can differential this equation with $\frac{\beta}{\mu}$ like \cite{Huang:2019zph}:
\begin{equation}
\begin{split}
    \frac{dA}{d(\frac{\beta}{\mu})}=\int_0^w\frac{8 \beta  \mu  z'(x)^2}{(1-z(x)) z(x)^2 \left(z(x) \left(z(x) \left(2 \beta ^2+\mu ^2
    z(x)-4\right)-4\right)-4\right)^2}\\
    \frac{1}{\sqrt{\frac{\frac{4 z'(x)^2}{(z(x)-1) \left(z(x) \left(z(x) \left(2 \beta ^2+\mu ^2
    z(x)-4\right)-4\right)-4\right)}+1}{z(x)^4}}}dx
\end{split}
\end{equation}
This will always be a positive value no matter what the z(x) is, which means the HEE will always increase as the momentum relaxation effect get stronger. Next we put this result back into MI: $d(MI)/d(\frac{\beta}{\mu})=d(S_A)/d(\frac{\beta}{\mu})+d(S_B)/d(\frac{\beta}{\mu})-d(S_{AB})/d(\frac{\beta}{\mu})$, we need to figure out which minimum surface dominate the MI behavior.

See Fig~\ref{fig:entanglement wedge}, the minimum surface near the boundary will canceled out with each other, so it is fair to say that the minimum surface that dominate this phenomenon is the minimum surface near the horizon, or $C_{abc}$, so $d(MI)/d(\frac{\beta}{\mu})\backsim-d(S_{AB})/d(\frac{\beta}{\mu})$, which will be a negative number.

\section{Conclusion}
\label{conclusion}

\qquad We established the entanglement of purification and the mutual information in a holographic relaxed superconductor system. To test our method, we took the configuration limit to the symmetry case, which has an analytic solution. We found out our method could approach to this analytic answer. We also check the inequality relation between the EoP and the MI/2, and the EoP will always bigger than the MI/2, which satisfied the relation. After confirmed our method, we discussed the EoP behavior in the high-temperature situation, we then found out it will approach to the straight line situation. We claimed that this is due to the fact that as the temperature increase, the geodesic will getting closer to the black hole horizon, and the distinct between the straight line solution and the actual minimum cross-section will become smaller. Next we focused on the dependence of the (a,b,c) configuration. We found out the EoP and the MI dependence on configuration are similar. When  "a" and "c" increase, both EoP and MI would increase, which in "c" case this was actually another inequality that EoP should satisfy. But when "b" increase, both EoP and MI would decrease.

In the last part of our research, we discussed how the momentum relaxation affect the EoP and MI. We found that the behavior of EoP and MI would be opposite as the momentum relaxation become stronger. This result is similar to the one in \cite{Huang:2019zph} for the small configuration. This told us that the EoP and MI still stand for different information in the entanglement theory. We separated the EoP part into two part to discuss and found that the background is the main factor that dominate this phenomenon. The MI part we return to the same conclusion in \cite{Huang:2019zph} that indeed it was dominated by the minimum surface near the horizon. We also calculated the critical "c" with fixed "a","b" in different $\beta/\mu$ situation, and it matched the behavior of MI. In all the situation above, we found our physical quantity like EoP, MI and $C_c$ would all appear a kink point at the critical temperature, which matched with the conclusion in \cite{Liu:2020blk}, the property of diagnosing the thermal transition will last in our model.

Our prominent result is that the superconductor effect will slightly change the behavior the momentum relaxation do to EoP and MI. The superconductor effect will increase the momentum relaxation effect do to EoP, but decrease the effect do to MI, at least in small configuration. This is a new phenomenon that we have never seen before.

In our research the superconductor effect really change how the momentum relaxation affect the information quantities. Unfortunately, it is hard to explain the mechanism between these complicated system. Also, there is a special kind of superconductor that establish in $\mu=0$ situation in the relaxed superconductor system, which we believe might have interesting information quantities. We will leave these to the future work.

\end{document}